\documentclass[10pt]{article}

\usepackage{fullpage}
\usepackage{setspace}
\usepackage{parskip}
\usepackage{titlesec}
\usepackage{xcolor}
\usepackage{lineno}
\usepackage[export]{adjustbox}
\usepackage{soul}

\usepackage[version=4]{mhchem}
\usepackage{siunitx}
\DeclareSIUnit\Molar{M}

\usepackage{mathtools}

\PassOptionsToPackage{hyphens}{url}
\usepackage[colorlinks = true,
            linkcolor = blue,
            urlcolor  = blue,
            citecolor = blue,
            anchorcolor = blue]{hyperref}
\usepackage{etoolbox}

\renewenvironment{abstract}
  {{\bfseries\noindent{\abstractname}\par\nobreak}\normalsize}
  {\bigskip}

\titlespacing{\section}{0pt}{*3}{*1}
\titlespacing{\subsection}{0pt}{*2}{*0.5}
\titlespacing{\subsubsection}{0pt}{*1.5}{0pt}

\usepackage{authblk}

\usepackage{graphicx}
\usepackage[space]{grffile}
\usepackage{latexsym}
\usepackage{textcomp}
\usepackage{longtable}
\usepackage{tabulary}
\usepackage{booktabs,array,multirow}
\usepackage{amsfonts,amsmath,amssymb}
\providecommand\citet{\cite}
\providecommand\cite{\cite}

\newif\iflatexml\latexmlfalse

\AtBeginDocument{\DeclareGraphicsExtensions{.pdf,.PDF,.eps,.EPS,.png,.PNG,.tif,.TIF,.jpg,.JPG,.jpeg,.JPEG}}

\usepackage[utf8]{inputenc}
\usepackage[english]{babel}

\newcommand{\fucone}{FUCONE} 
\newcommand{\vectorize}[1]{\ensuremath{\text{vec}(#1)}}
\newcommand{\fcidx}{\ensuremath{c}} 
\newcommand{\diversity}{\ensuremath{\Theta}}
\newcommand{\imcoh}{ImCoh}
\newcommand{\plv}{PLV}
\newcommand{\pli}{PLI}
\newcommand{\wpli}{wPLI2-d}
\newcommand{\aec}{AEC}
\newcommand{\cov}{Cov+EN}
\newcommand{\instant}{Instantaneous}

\providecommand{\keywords}[1]{\textbf{\textit{Keywords ---}} #1}

\begin{document}

\doublespacing

\title{Functional connectivity ensemble method to enhance BCI performance (FUCONE)\footnote{This work has been submitted to the IEEE for possible publication. Copyright may be transferred without notice, after which this version may no longer be accessible.}}

\author[a*]{Marie-Constance Corsi}
\author[b]{Sylvain Chevallier} 
\author[a]{Fabrizio De Vico Fallani}
\author[c]{Florian Yger}

\affil[a]{Sorbonne Université, Institut du Cerveau - Paris Brain Institute - ICM, CNRS, Inria, Inserm, AP-HP, Hôpital de la Pitié Salpêtrière, F-75013, Paris, France}
\affil[b]{LISV, UVSQ, Université Paris-Saclay, France}
\affil[c]{LAMSADE, PSL, Université Paris Dauphine, France}
\affil[*]{Corresponding author: 
Marie-Constance Corsi \url{marie.constance.corsi@gmail.com}}

\vspace{-1em}
\begingroup
\let\center\flushleft
\let\endcenter\endflushleft
\maketitle
\endgroup

\selectlanguage{english}

\newpage
\begin{abstract}
Objective: Relying on the idea that functional connectivity provides important insights on the underlying dynamic of neuronal interactions, we propose a novel framework that combines functional connectivity estimators and covariance-based pipelines to improve the classification of mental states, such as motor imagery. Methods: A Riemannian classifier is trained for each estimator and an ensemble classifier combines the decisions in each feature space. 
A thorough assessment of the functional connectivity estimators is provided and the best performing pipeline among those tested, called \fucone{}, is evaluated on different conditions and datasets. Results: Using a meta-analysis to aggregate results across datasets, \fucone{} performed significantly better than all state-of-the-art methods. Conclusion: The performance gain is mostly imputable to the improved diversity of the feature spaces, increasing the robustness of the ensemble classifier with respect to the inter- and intra-subject variability. Significance: Our results offer new insights into the need to consider functional connectivity-based methods to improve the BCI performance.
\end{abstract}


\section{Introduction}
\label{sec:introduction}
Brain-computer interfaces (BCIs) aim at translating brain activity into commands for control or communication~\cite{wolpaw_braincomputer_2002}. This technology raises hope for patients as it presents many clinical applications from the control of wheelchairs~\cite{carlson_brain-controlled_2013} to the communication with relatives \cite{naci_brains_2013} going through stroke rehabilitation~\cite{cervera_braincomputer_2018}.

Mastering BCI control via a voluntary modulation of the cerebral activity remains a learned skill that requires subject training. Besides, a non-negligible portion of BCI users (around 30 \%) cannot control a BCI system after completing a training program. This phenomenon, associated to a high inter-subject variability, is referred in the literature as the ``BCI inefficiency''~\cite{thompson_critiquing_2018} and affects the usability of the BCI among the patients. To circumvent it, several approaches have been adopted. Among them are the proposal of new interaction paradigms~\cite{vidaurre_enhancing_2019},
the search of BCI performance predictors~\cite{blankertz_neurophysiological_2010-1, ahn_performance_2015, hammer_psychological_2012} and the design of more sophisticated classification algorithms to better discriminate the subjects' mental state~\cite{dahne_spoc_2014,de_cheveigne_joint_2014,sabbagh_predictive_2020, edelman_noninvasive_2019, suma_spatial-temporal_2020}. In particular, Riemannian geometry-based methods \cite{yger2016riemannian, congedo2017riemannian} are now the gold standard by reaching the state-of-the-art performance~\cite{lotte_review_2018} and by winning several data competitions\footnote{See for example the $6$ competitions won by A. Barachant: \url{http://alexandre.barachant.org/challenges/} (accessed on December, 17 2021)}~\cite{corsi_riemannian_2021}.

Another approach would consist in taking into account the subjects' specificity by considering alternative features to be classified. In the particular case of left (or right) motor-imagery (MI) based BCI, the system relies on the desynchronization effect associated with a decrease of the power spectra computed within the contralateral sensorimotor area~\cite{pfurtscheller_event-related_1999}. Therefore, experimenters typically choose power spectra estimated in channels located above sensorimotor area as features.  However, it does not capture the changes in communication between brain areas or said differently, it does not take into account the interconnected nature of brain activity.
Functional Connectivity (FC), estimating the interaction between different brain areas~\cite{de2014graph,bastos_tutorial_2016}, is a promising tool for BCI~\cite{hamedi_electroencephalographic_2016, gonzalez-astudillo_network-based_2020}. Indeed, it has been proved to provide alternative features to discriminate subjects' mental states~\cite{gysels_phase_2004, billinger_single-trial_2013, hamedi_electroencephalographic_2016, feng_functional_2020, cattai_phaseamplitude_2021,cattai_improving_2021} and to study brain networks reorganization underlying MI-based BCI training~\cite{corsi_functional_2020, corsi_bci_2021}. 

On the above-mentioned elements, we hypothesized that combining functional connectivity estimators, Riemannian geometry and ensemble learning will lead to a performance improvement. Furthermore, we expect that the analysis of the selected features will give insights on the brain interactions the most relevant during MI-based BCI performance. 

Promising preliminary results have been obtained during the IEEE WCCI Clinical BCI Challenge~\cite{corsi_riemannian_2021}, where this approach ranked first on one of the two proposed tasks. In this paper, we led a comprehensive study in order to optimize the key parameters (functional connectivity estimators and frequency bands of interest) and to ensure the replicability of our results.

\begin{table*}[t]
\caption{Functional connectivity metrics considered in the study}
\label{tab:tab1}
\center
\begin{tabular}{lllll}
FC metric & Abbreviation & Category & Ref. \\
\hline
Instantaneous coherence & \instant &    Spectral coherence   &      \cite{pascual-marqui_instantaneous_2007} \\
Imaginary Coherence & \imcoh{} &   Spectral coherence      &      \cite{nolte_identifying_2004} \\
Phase-Locking Value & \plv{} &   Phase estimation           &     \cite{lachaux_measuring_1999} \\
Phase-Lagged Index & \pli{} &   Phase estimation           &    \cite{stam_phase_2007}  \\
debiased estimator of squared wPLI & \wpli{} &  Phase estimation      &  \cite{vinck_improved_2011}    \\   
Amplitude Envelope Coupling & \aec{} &  Amplitude coupling         &  \cite{hipp_large-scale_2012,brookes_measuring_2011}    \\ 
\hline
\end{tabular}
\end{table*}

\section{Materials and Methods}

\subsection{Riemannian geometry}

Riemannian geometry deals with smoothly
curved spaces that locally behave like Euclidean spaces. For example, orthogonality or definite positive constraints in matrices can be shown to be smooth and can be dealt with a Riemannian geometry. In the manner of Russian dolls, Riemannian manifolds are built as the result of nested mathematical structures and this description goes beyond the scope of this introduction, we refer any interested reader to~\cite{yger2016riemannian} and references therein. The key point to remember is that a Riemannian manifold can be locally linearized at any point by a tangent space and each tangent space can be equipped with a scalar product.   
The notion of tangent space at a given point can be understood as the gradient of the curves passing through this point. 

For any Riemannian manifold, there exists a pair of inverse operations mapping points from the manifold to any given tangent space and vice versa. The exponential mapping  can transport any point from the tangent space to the manifold. Conversely, the logarithmic mapping can transport any point from the manifold to the tangent space defined at a given reference point\footnote{Note that each tangent space is defined locally and mapping a point far away from the reference point implies some heavy distortion.}. Then, the scalar product that equips each tangent space enables to measure distances in any tangent space. The definition of the geodesic, \emph{i.e.}, the shortest path on the manifold between two points, is inherited from the choice of the scalar product. For certain matrix constraints, several geometries can be defined over the same space but the choice of equipped scalar product will make a difference. The length of such geodesic then serves to measure the distance between two points along the curved surface. 

As a consequence, it becomes possible to extend some approaches in the Riemannian domain by substituting the Euclidean
distance by a Riemannian distance. For example, for a given manifold $\mathcal{M}$ equipped with a Riemannian distance, the Fr\'echet mean extends the concept of mean and is defined as :
\begin{align*}
    \Bar{X} = arg \min_{X\in \mathcal{M}} \sum_i \delta_r^2(X_i, X) 
\end{align*}

Note that this is an optimization problem on a matrix manifold and taking advantage of the tangent spaces (together with approximations of the exponential mapping) it becomes possible to efficiently solve such a problem~\cite{absil2009optimization, boumal2020introduction}.
A Riemannian average can be used to build a simpler yet robust classifier known as Minimum Distance to the Mean (MDM) where each class is represented by its Riemannian average and matrices are classified according to the closed class. 

Let us focus on the space of SPD matrices, noted $\mathcal{P}_n = \{X \in \mathbf{R}^{n \times n} | X = X^\top , X \succ 0\}$. In other words, it is the space of symmetric matrices of strictly positive eigenvalues. It has been used successfully for handling covariance matrices from EEG signals. At every point $X$ of the space of
SPD matrices, the tangent space $T_X \mathcal{P}_n $. Depending on
the choice of the scalar product to equip the tangent spaces,
one Euclidean and two different Riemannian geometries are defined for any pair of SPD matrices $X$ and $Y$ 
\begin{itemize}
    \item Euclidean distance : 
        $\delta_e \left(X,Y\right) = ||X-Y||_\mathcal{F}$
    \item LogEuclidean : $
        \delta_{le}\left(X,Y\right) = ||\log(X)-\log(Y)||_\mathcal{F}$
    \item Affine Invariant : $
        \delta_{r}\left(X,Y\right) =||\log (X^{-\frac{1}{2}}YX^{-\frac{1}{2}})||$
    \end{itemize}

The Euclidean distance is obtained by considering the space of SPD matrices as a subspace of the Euclidean space of symmetric matrices. Although this is quite natural and easy to implement, the Euclidean geometry suffers from three major drawbacks as documented in~\cite{yger2016riemannian}. First of all, in this geometry, it is possible to interpolate between SPD matrices but extrapolation can lead to a non-SPD result. Then, it is affected by the swelling effect, as the Euclidean interpolation between two matrices can have a bigger determinant as each of them and if we consider the determinant as a measure of information, interpolation then creates some artifact. Finally, it is not invariant to affine transformations (such as the left and right multiplication by an invertible matrix) and for example, if we scale $X$ and $Y$ by scalar bigger than $1$, the distance between them will grow. In the context of EEG signal processing, such linear transformations can occur when the electrodes location change slightly over time or when the recorded signals result from different sources of activity mixed together. Therefore, being sensitive to affine transformation can lead to misclassifications. On the contrary, the AIRM distance is immune to those drawbacks but its computation is more computationally demanding. 
Hence, the LogEuclidean distance has been proposed as a trade-off between AIRM and LogEuclidean distances, retaining some interesting properties of the AIRM distance while being faster to compute~\cite{chevallier_review_2021}. Overall, the use of a Riemanning geometry (LogEuclidean or AIRM) leads to a simpler feature extraction step tampering with the need for spatial filtering and a less complex data processing pipeline.

So far, the Riemannian geometry has been successfully applied for manipulating covariance matrices from EEG signals. The diagonal of those matrices carries information related to the electrodes power and it relates to the kind of information extracted by Common Spatial Patterns (CSP)-based methods. However, those matrices also contain the electrodes covariances and it is then a richer feature in terms of carried out information. Covariance matrices are at the heart of the interest for Riemannian geometry for BCI. However, connectivity information can be extracted in the form of SPD matrices and contains complementary information compared to covariance matrices.

\subsection{Functional connectivity for EEG signals}
\label{FC}

Functional connectivity (FC) aims at estimating the interaction between different brain areas ~\cite{de2014graph,bastos_tutorial_2016}. There is plethora of FC estimators. Here, we considered complementary undirected FC estimators to identify which of them, being symmetric and positive definite, combined with Riemannian geometry, would best classify the data. For a given FC estimator, we averaged the FC values within specific frequency bands of interest. To ensure that all resulting matrices were SPD, we applied an adaptive regularization to make semi-definite positive matrices to SPD matrices that consisted of adding the identity matrix multiplied by small positive constant.

\subsubsection{Spectral estimators} 
Two spectral estimators were considered here: the imaginary coherence and the instantaneous coherence. Both of them result from the coherency, \emph{i.e.}, the normalized cross-spectral density obtained from two given signals:

\begin{equation}
\mathrm{Coh}_{ij}(f)=\frac{S_{ij}(f)}{\sqrt{S_{ii}(f).S_{jj}(f)}}
\end{equation}
where $S_{ij}(f)$ the cross-spectral density and $S_{ii}(f)$ the auto-spectral density.

The imaginary part of coherence (\imcoh) is less sensitive to signal leakage and volume conduction effects~\cite{nolte_identifying_2004,colclough_how_2016}.
The instantaneous coherence (\instant) corresponds to the real part of the coherency~\cite{pascual-marqui_instantaneous_2007}.

We estimated the cross-spectral density of each pair of EEG channels time courses, using multitapers \cite{slepian_prolate_1978}, with time windows of length 1 s with an overlap of 0.5 s for each trial.

\subsubsection{Phase estimators} 
The Phase-Locking Value (\plv), and the Phase-Lag Index (\pli) assesses phase synchrony between signals in a specific frequency band.
The \plv{} corresponds to the absolute value of the mean phase between $s_{i}$ and $s_{j}$, is defined as follows~\cite{lachaux_measuring_1999,tass_detection_1998, aydore_note_2013}:
\begin{equation}
\plv =|e^{i\Delta\phi(t)}|
\end{equation}
where $\Delta\phi(t)=arg(\frac{z_i(t).z_j^{*}(t)}{|z_i(t)|.|z_j(t)|})$. 
$\Delta\phi(t)$ represents the associated relative phase computed between signals and $z(t)=s(t)+i.h(s(t))$ the analytic signal obtained from the signal $s(t)$.

The \pli{} assesses the asymmetry of the phase difference \cite{stam_phase_2007}. As a result, it is less sensitive to shared signals at zero phase lag.
\begin{equation}
\pli =|\text{sign} \sin\Delta\phi|
\end{equation}

Given its sensitivity to volume conduction and noise, we considered an additional phase-based metrics resulting from the \pli{}: the debiased estimator of squared weighted Phase Lag Index \cite{vinck_improved_2011} (\wpli). The \wpli{} is notably less sensitive to uncorrelated noise sources and its unbiased version enables a reduction of the sample-size bias.
Functional connectivity estimations were made by using multitapers \cite{slepian_prolate_1978}, with time windows of length 1 s with an overlap of 0.5 s for each trial.

\subsubsection{Amplitude coupling estimator} 
We computed the Amplitude Envelope Correlation (\aec) \cite{ hipp_large-scale_2012, brookes_measuring_2011,colclough_how_2016} which relies on the linear correlations of the envelopes of the band-pass filtered signals obtained from Hilbert transform. 

All the metrics considered in this paper are summarized in Table~\ref{tab:tab1}. They were estimated using MNE~\cite{gramfort_mne_2014} and PyRiemann\footnote{\url{https://pyriemann.readthedocs.io} (accessed on December, 17 2021)} toolboxes.
Once a given functional connectivity estimator $FC$ is computed, the associated dataset is projected onto the SPD manifold to get the SPD matrix $X_{FC}$ that will be used for the Riemannian operation.

\subsection{Proposed approach: FUnctional COnnectivity eNsemble mEthod (\fucone{})} 

\begin{figure*}
\begin{center}
		\includegraphics[width=0.9\textwidth]{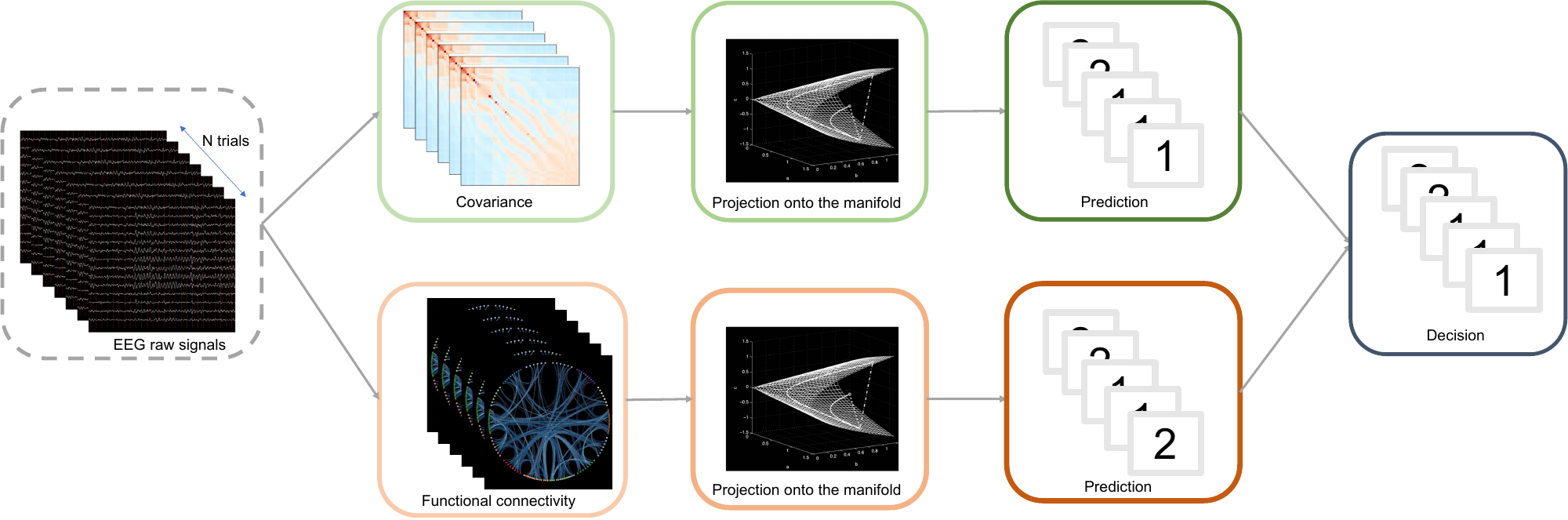}
		\caption{\fucone{} approach (generic view).
		{\label{519840}}
		}
\end{center}
\end{figure*}

Our approach consists of combining FC estimators and Riemannian Geometry with two levels of classification, as shown on Figure~\ref{519840}. 
For each type of estimator considered (\emph{i.e.}, covariance and functional connectivity metrics), we performed a first-level classification in the tangent space using an Elastic-Net classifier via scikit~\cite{scikit-learn, sklearn_api}. This classifier aimed to find the weights $w$ that minimized the following objective function:
\begin{equation}
\min_w || \vectorize{X}w - y ||^2_2 + \alpha \rho ||w||_1 + \alpha(1 - \rho)||w||^2_2
\end{equation}
The hyperparameters were set up through a preliminary study, with $\alpha = 1$ and $\rho = 0.15$, meaning that the penalty was a combination of $\ell_1$ and $\ell_2$ norms. 

The first-level classifiers were used as a stacked ensemble of classifiers, using an Elastic-Net as an ensemble classifier, with $\alpha = 1$ and $\rho = 0.15$. This ensemble classifier was trained through a stratified $k$-fold that contained five splits.

\subsection{Dataset for preliminary study} 
To identify the most suited parameters, we worked with a publicly available dataset~\cite{cho_eeg_2017}. The authors performed a BCI experiment of left and right hands motor imagery with a 64-channel EEG montage. The sessions were composed of five to six runs each. For more details on the dataset, the reader can refer to the description provided by Cho et al \cite{cho_eeg_2017}.
For a sake of efficiency and to get a subset of representative subjects, we preselected the five most and the five least responsive subjects, based on the classification accuracy obtained at the individual level. For that purpose, we based our selection on the FgMDM classification scores, which corresponds in our case to the state-of-the-art classification algorithm in motor imagery-based BCI  \cite{barachant2010riemannian, barachant_classification_2013}. We preselected the subjects 14, 43, 50, 35 \& 3 (as the most responsive subjects) and 38, 40, 17, 7 \& 29 (as the least responsive subjects). For a given FC estimator, a time window of [0s, 3s] was considered. 

To verify the generalization property of our approach, we conducted a replicability analysis on several datasets. The details are provided in section~\ref{sec:repli}.

\subsection{Performance assessment}
\subsubsection{Baseline pipelines}
Our approach was compared to the state-of-the-art algorithms in BCI. For this purpose, we considered the following pipelines:
\begin{itemize}
    \item "RegCSP+shLDA" refers to a CSP based on Ledoit-Wolf shrinkage of the covariance term, followed by a Linear Discriminant Analysis (LDA) classifier~\cite{lotte_regularizing_2011}
    \item "CSP+optSVM" corresponds to a CSP followed by a Support Vector Machine (SVM) with grid search to find optimal parameters~\cite{ang_filter_2008}
    \item "FgMDM": it refers to the geodesic filtering  followed by a Minimum Distance to Mean classification~\cite{yger2016riemannian}
\end{itemize}

The classification scores for all pipelines, \fucone{} included, were evaluated with a balanced accuracy measure, either using 5-fold cross-validation for within-session evaluation or with \emph{leave-one-session-out} for cross-session evaluation.

\subsubsection{Statistical analysis}
Within a dataset, depending on whether the number of subjects exceeded 20, a one-tailed paired $t$-test or a Wilcoxon signed-rank test for each pair of pipelines was considered. To assess the replicability of our method, we tested it on several datasets. In our study, we used two values, the standardized mean difference (SMD) and the $p$-value. The SMD value is a normalized assessment of the difference between the means of the classification performance between two tested pipelines. The $p$-value corresponds to the probability of obtaining a significant difference between the classification performance obtained from these two pipelines. In accordance with the approach proposed in \cite{jayaram_moabb_2018}, we obtained this statistical difference via the Stouffer’s method \cite{stouffer_american_1949} that combines p-values resulting from the Wilcoxon signed-rank test on each dataset taken separately. Then a Bonferroni correction was performed to prevent it from false positive. 

\subsubsection{Diversity}
To determine the possible improvement brought by a pipeline relying on a given functional connectivity estimator $\fcidx$, to covariance-based pipelines $\text{Cov}$, we built a diversity index upon the work of~\cite{kuncheva2003measures}.
Diversity corresponds here to the proportion of trials misclassified by the $\text{Cov}$ that have actually been correctly classified by $\fcidx$. 
Extending the definition given in~\cite{chevallier_extending_2020}, the diversity $\diversity$ is defined as follows:
\begin{equation}
    \diversity = \Omega^{\text{Cov}}_{FN+FP} \cap \Omega^{\fcidx}_{TP+TN}
\end{equation}
with $\Omega^{\text{Cov}}_{FN+FP}$ the number of trials incorrectly classified with covariance-based pipeline, that is false positive $FP$ and false negative $FN$ trials. The number of trials correctly classified by $\fcidx$ pipeline $\Omega^{\fcidx}_{TP+TN}$, that is true positive $TP$ and true negative $TN$. We distinguished two kinds of diversity measurements: the first one is the \textit{potential improvement} $\frac{\diversity}{|\Omega|}$, with $|\Omega|$ is the total number of trials and the \textit{relative diversity} $\frac{\diversity}{\Omega^{\text{Cov}}_{FN+FP}}$. 

\section{Results} 
\subsection{Thorough evaluation of FC-based pipelines}
In this section, we investigated the relative contributions of the key
parameters for FC-based pipelines, with the objective of selecting the best parameters for designing the \fucone{} approach. Among the considered items were the FC estimators, the frequency bands of interest and the combination of first-level classifiers into our workflow.

\subsubsection{Impact of FC estimators}
\begin{figure*}
\begin{center}
		\includegraphics[width=\textwidth]{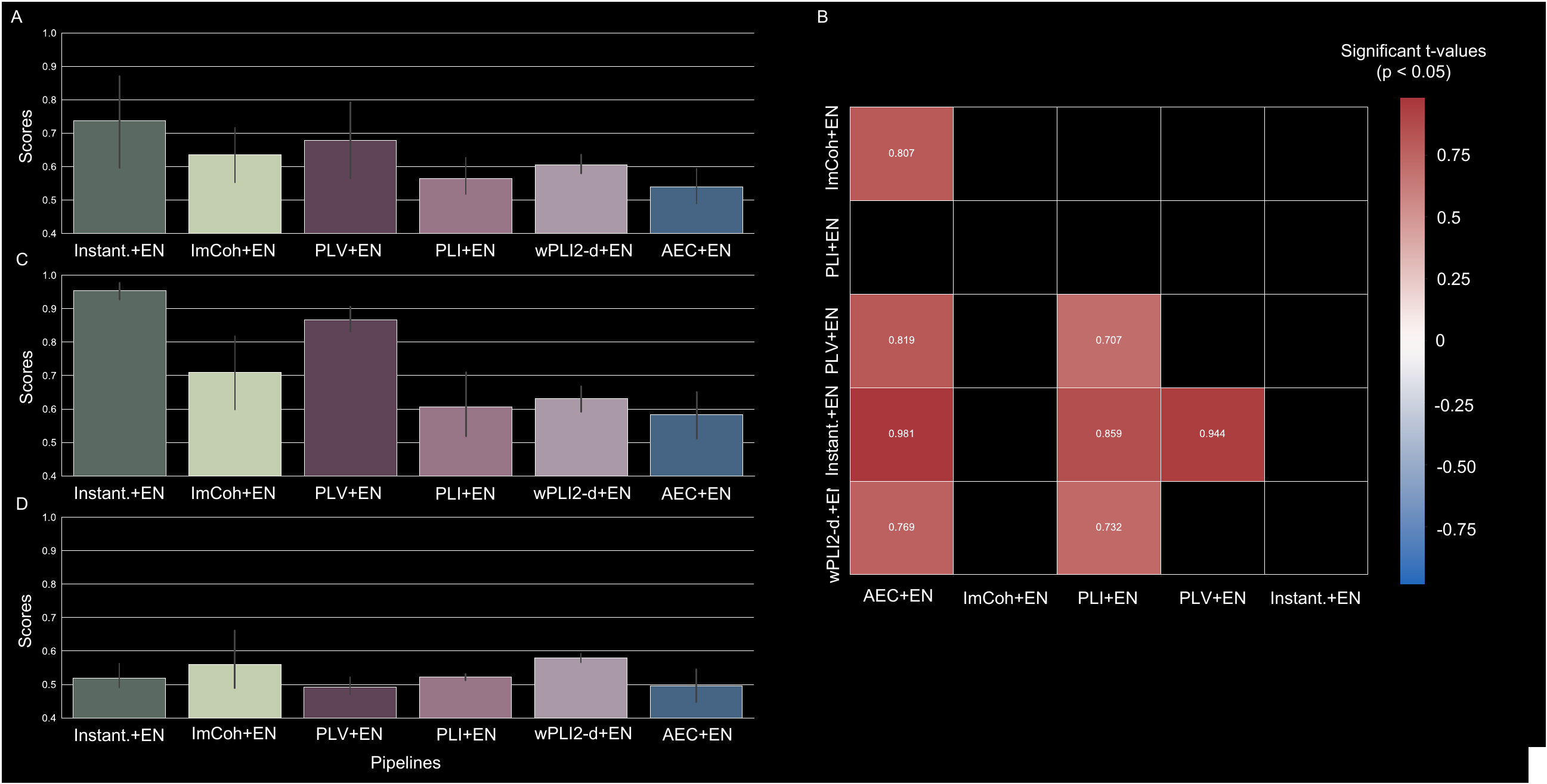}
		\caption{Functional connectivity estimators - (A) Group-level analysis. Barplots of the accuracy obtained over the group of ten subjects. (B) Group-level analysis. Statistical comparisons of each couple of pipelines. The color represents the significance level of the difference of accuracy, in terms of t-values, obtained with two given pipelines associated to two specific FC estimators. We plotted only the significant interactions ($p < 0.05$).
		(C) Subgroup analysis. Barplots of the obtained accuracy over the group of most responsive subjects ($N$=5 subjects). (D) Subgroup analysis. Barplots of the obtained accuracy over the group of least responsive subjects ($N$=5 subjects).
		{\label{519843}}
		}
\end{center}
\end{figure*}

The first step consisted of identifying the FC estimators that discriminated the best signals. We considered the estimators defined in section~\ref{FC}. We compared the accuracy obtained across the ten subjects (see Figure~{\ref{519843}}A). Here, we considered the 8-35~Hz frequency band. We observed that the three best FC estimators were the instantaneous coherence (0.74 $\pm$ 0.23), the phase-locking value (0.68$\pm$0.20) and the imaginary part of coherence (0.63$\pm$0.15). Indeed, they performed significantly better than \aec{} ($p <$ 0.017). \instant{} and \plv{} performed significantly better ($p <$ 0.03) than \pli{}, and \wpli{} (see Figure~{\ref{519843}}B). 
Then, we explored the results at the sub-group level. We considered the most and the least responsive subjects separately in order to identify potential FC estimators susceptible to favor the most and/or the least responsive subjects (see Figure~{\ref{519843}}B). As expected, \instant{}, \plv{} and \imcoh{} showed the highest performance with the most responsive group, with a mean accuracy respectively of 0.94, 0.87 and 0.71 (see Figure~{\ref{519843}}C). 
\wpli{} and \imcoh{} showed higher accuracy than the other estimators with the least responsive group, with an accuracy respectively of 0.58 and 0.56 (see Figure~{\ref{519843}}D). 
In order to take into account both the group tendency and the need to provide alternative methods to discriminate least responsive subjects' mental state, we reduced our pool of FC estimators to the following estimators: \instant, \plv{}, \imcoh{} and \wpli{}. 

\subsubsection{Frequency bands of interest}
\begin{figure*}
\begin{center}
		\includegraphics[width=0.81\textwidth]{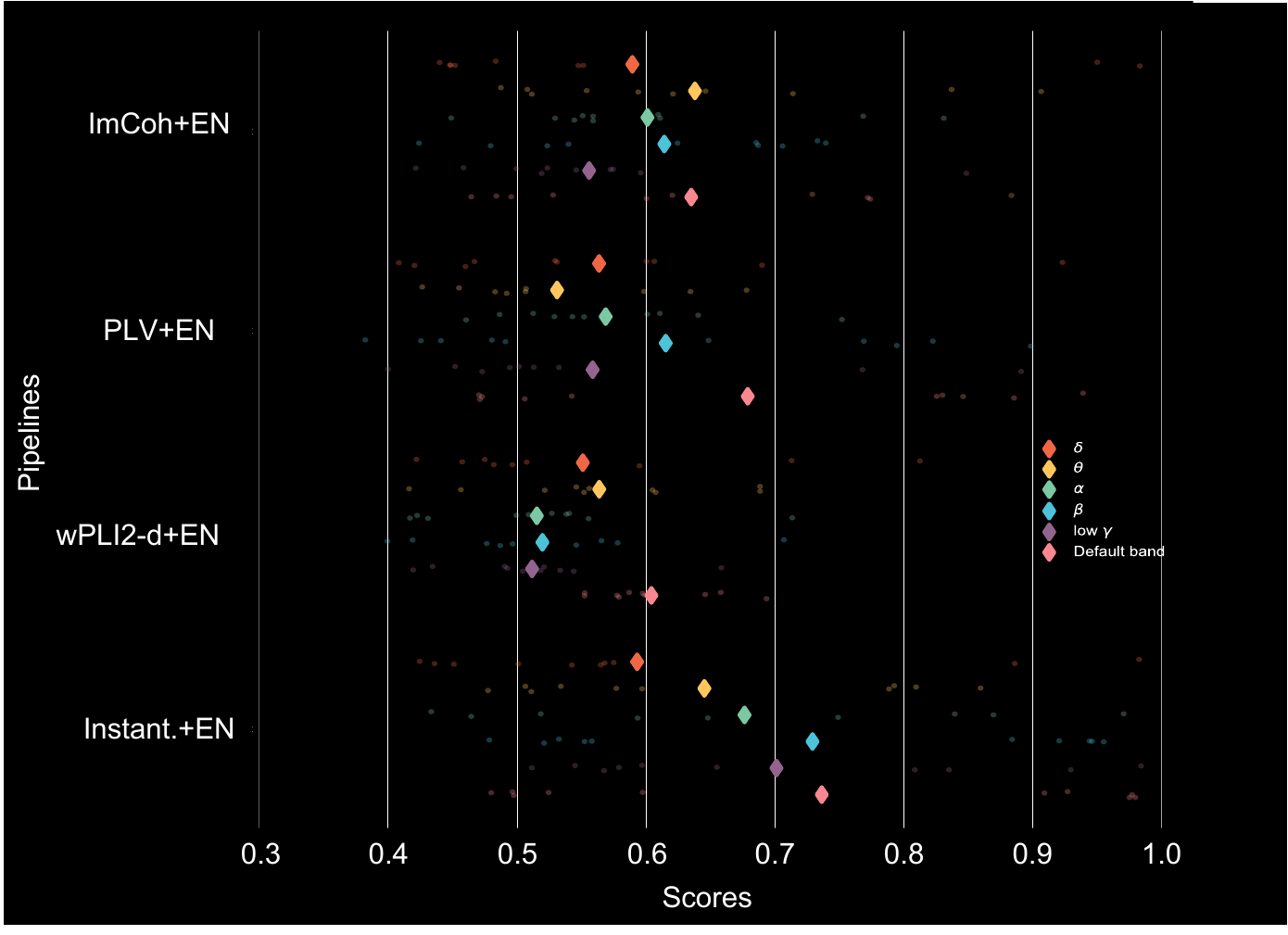}
		\caption{Frequency bands - Group-level analysis. For each pipeline, we plotted the distribution of the scores obtained over the subjects. Each color corresponds to a specific frequency band.
		{\label{519846}}
		}
\end{center}
\end{figure*}

To further investigate the influence of the key parameters on the performance of our approach, we compared the accuracy over different frequency bands. The sensorimotor rhythms gather mu-rhythm (\emph{i.e.}, between 8 and 12~Hz), often accompanied with a beta component (\emph{i.e.} 12 to 30 Hz) and a low gamma component (between 30 and 40Hz)~\cite{kubler_chapter_2016}. Here, we broadened our study to the following frequency bands: delta $\delta$ [2, 4~Hz], theta $\theta$ [4, 8~Hz], alpha $\alpha$ [8, 12~Hz], beta $\beta$ [15, 30~Hz], (low) gamma $\gamma$ [30, 45~Hz] and the default band [8, 35~Hz]. We performed the classification for each preselected FC estimator and each frequency band of interest separately. We analyzed the performance obtained (see Figure~\ref{519846}). We did not observe a frequency band effect (one-way ANOVA, $p >$ 0.05), meaning that we did not observe that a specific frequency band outperformed. Nevertheless, it appeared that higher scores were obtained in the default band since we obtained an average accuracy of 0.66$\pm$0.17, whereas we obtained 0.57$\pm$0.17, 0.59$\pm$0.13, 0.59$\pm$0.13, and 0.62$\pm$0.17, respectively for the delta, the theta, the alpha, the beta, and the gamma bands. Furthermore, as previously mentioned, the default band is the closest to the sensorimotor rhythm. As a result, we chose to work with this frequency band.

\subsubsection{Stacking classifiers for ensemble learning}

\begin{figure*}
\begin{center}
		\includegraphics[width=0.9\textwidth]{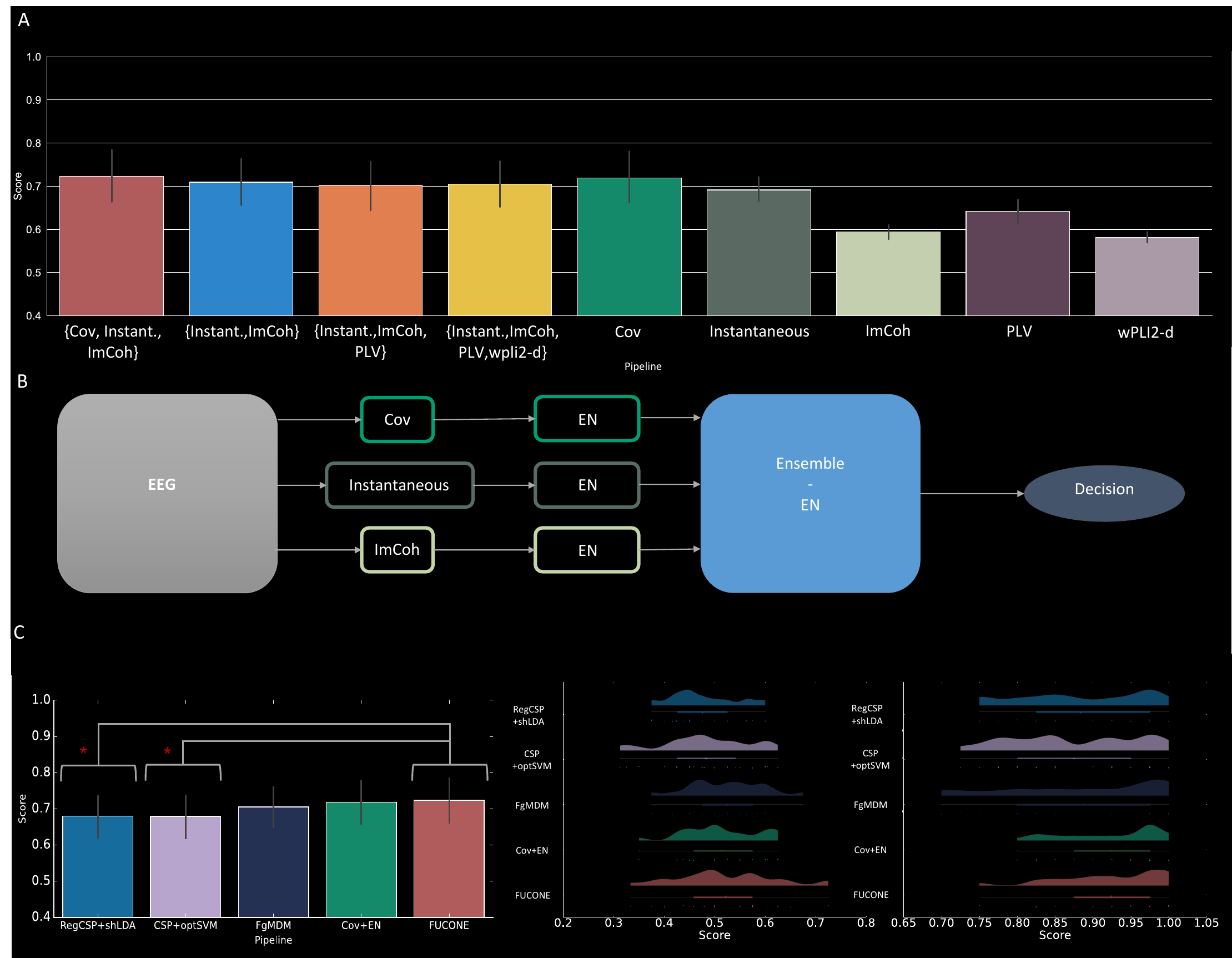}
		\caption{Stacking classifiers for ensemble learning. (A) Performance comparison between different combinations of pipelines and each FC estimator taken separately. Each bar corresponds to a specific configuration.
		(B) \fucone{} approach. From EEG recordings, covariance, \instant{} and \imcoh{} estimations are computed and classified with an Elastic-Net (EN) approach followed by an ensemble classifier relying on an Elastic-Net that provides the decision.
		(C) Comparison with the state-of-the-art. On the left, the bar plots represent the scores obtained from each pipeline at the group-level (* means $p < 0.05$). The last two raincloud plots represent, for each pipeline, the distribution of the scores over the subgroups composed by respectively the most and the least responsive subjects.
		{\label{519851}}%
		}
\end{center}
\end{figure*}

Once the preselected FC estimators and the frequency band identified, the next step consisted of finding the best combination of single-level classifiers. For that purpose, we considered different combinations and compared them to the pipelines associated with the FC estimators taken separately (see Figure~{\ref{519851}A}). 

At the group level, the configuration that performed the best consisted of combining the covariance, the instantaneous coherence and the imaginary part of coherence estimations with an average score of 0.72$\pm$0.22. It was the only configuration that presented a higher performance than both FC estimators and covariance. The three other combinations, (\instant + \imcoh), (\instant + \imcoh + \plv), (\instant + \imcoh + \plv + \wpli), 
showed similar performance, always superior to the ones obtained with FC estimators taken alone, with respectively 0.71$\pm$0.20, 0.70$\pm$0.20 and 0.70$\pm$0.20.
As a result, we chose to work with the configuration presented in Figure~\ref{519851}B that includes the estimation of the covariance, the estimation of the instantaneous coherence and the imaginary part of coherence. All of them were associated with an Elastic-Net (EN) classifier. The ensemble classifier was also an Elastic-Net. In the next sections, when we mention our pipeline \fucone{} we will refer to the configuration proposed in Figure~\ref{519851}B.

We compared the performance of our \fucone{} pipeline with the state-of-the-art (see Figure \ref{519851}C). 
At the group level, we observed that our approach gave the highest score with respect to the Cov+EN, the FgMDM, the CSP+optSVM, and the RegCSP+shLDA approaches that showed respectively of 0.71$\pm$0.22, 0.70$\pm$0.20, 0.68$\pm$0.22 and 0.68$\pm$0.20. Notably, our approach performed better both for the most and the least responsive subjects.

\subsection{\fucone{}, a replicable approach to improve MI classification}  
\label{sec:repli}
In order to assess the robustness and the replicability of our approach, we tested it through a larger number of subjects, datasets and motor imagery tasks. We also considered two types of performance evaluation: within-session and cross-session.

\subsubsection{Considered datasets}
To perform this analysis, we only used open-access datasets in which the participants are healthy. The Table~\ref{tab:tab2} provides a description of the datasets we used.

\begin{table*}[]
\center
\caption{Datasets included in the replicability study}
\label{tab:tab2}
\begin{tabular}{cccccccc}
ID & \#subjects & \#channels & Sampling rate & \#sessions & \#tasks  & \#trials/class  & Ref. \\
\hline \hline
Cho2017  & 52 & 64 & 512~Hz & 1  & 2  & 100 & \cite{cho_eeg_2017}  \\
\hline
Weibo2014 & 10 & 60 & 200~Hz & 1 & 7  & 80  & \cite{yi_evaluation_2014}  \\
\hline
Schirrmeister2017 & 14 & 128 & 500~Hz & 1 & 4  & 120 & \cite{schirrmeister_deep_2017}   \\
\hline
Zhou2016  & 4  & 14 & 250~Hz & 3  & 3  & 160 & \cite{zhou_fully_2016}   \\
\hline
001-2014 & 10 & 22 & 250~Hz & 2 & 4 & 144 & \cite{tangermann_review_2012}  \\
\hline
001-2015 & 13 & 13 & 512~Hz & 1 & 2 & 200 & \cite{faller_autocalibration_2012}  \\
\hline
002-2014  & 14 & 15 & 512~Hz & 5 & 2 & 80  & \cite{steyrl_random_2016}   \\
\hline
004-2014 & 10 & 3  & 250~Hz & 1 & 2 & 360 & \cite{leeb_brain-computer_2007} \\
\hline \hline
\end{tabular}
\end{table*}

\subsubsection{Within-session evaluation} 
In this section, we compared our score with those obtained with the state-of-the-art pipelines (see Figure~\ref{fig:meta}A) by applying a within-session evaluation. We pushed the limits of our approach by considering datasets with an increased number of tasks performed by the subjects. In particular, it enabled us to verify that our approach was not specific to the classic "left vs right" hand motor imagery.

\textit{2 classes - right hand versus feet (rf)} -- 
In this case, we considered five datasets. For four of them, \fucone{} showed the best results (see Figures \textit{S1-2}), both in terms of average accuracy (ranging from 0.82 to 0.91) and variability (ranging from$\pm$0.08 to$\pm$0.14). In Zhou2016, the pipeline \cov{} showed the best results (0.913$\pm$0.07) while ours was in second position (0.905$\pm$0.08).

\textit{2 classes - left hand versus right hand (lhrh)} --
Four datasets were considered here. \fucone{} presented the best results (see Figures \textit{S3-4}) for three of them (001-2014, 004-2014, and Schirrmeister2017) with an average accuracy ranging from 0.74 to 0.84 and a standard deviation ranging from$\pm$0.13 to$\pm$0.16. In Zhou2016, our approach ranked second (0.873$\pm$0.101) after the \cov{} pipeline (0.874$\pm$0.106).

\textit{3 classes} --
We took into account three datasets. For two of them (001-2014, Weibo2014), \fucone{} showed the best performance (see Figures \textit{S5-6}), with accuracy ranging from to 0.73 to 0.80 and standard deviations ranging from$\pm$0.13 to$\pm$0.14. Here again, in Zhou2016, the pipeline \cov{} showed the best results (0.844$\pm$0.08) while ours was in second position (0.839$\pm$0.08).

\textit{4 classes} --
Two datasets were considered here (Schirrmeister2017 and Weibo2014). In both cases, \fucone{} showed the best accuracies (see Figures \textit{S7-8}) (ranging from 0.80 to 0.86) and variability (ranging from$\pm$0.09 to$\pm$0.10).

\subsubsection{Cross-session evaluation}
To investigate the possibility to predict the performance across sessions for a given subject, we led a study with datasets that invited the participants to come twice. The training set gather all but one session, that is used a test set to estimate the prediction score, as in~\cite{khazem_minimizing_2021}. We compared our score with those obtained with the state-of-the-art pipelines (see Figure~\ref{fig:meta}B) by applying a cross-session evaluation this time. Here, we only had access to datasets associated with two motor imagery tasks (see Table \ref{tab:tab2}).

\textit{2 classes - lhrh} --
In this case, we tested the pipelines on three datasets (see Figures \textit{S9-10}). For two of them, \fucone{} showed the best performance (from 0.72$\pm$0.13 to 0.78$\pm$0.12). In Zhou2016, the pipeline \cov{} showed the best results (0.794$\pm$0.14) while ours was in second position (0.791$\pm$0.14).

\textit{2 classes - rh-f} --
In this case, three datasets were considered. For all of them, \fucone{} ranked second after the \cov{} one (see Figures \textit{S11-12}), with accuracy ranging from 0.79 to 0.86 (respectively from 0.79 to 0.85 with our pipeline) and standard deviations ranging from$\pm$0.09 to$\pm$0.14 (respectively from$\pm$0.10 to$\pm$0.13 with our pipeline).

\subsubsection{Meta-analysis}
\begin{figure*}
\begin{center}
		\includegraphics[width=450pt]{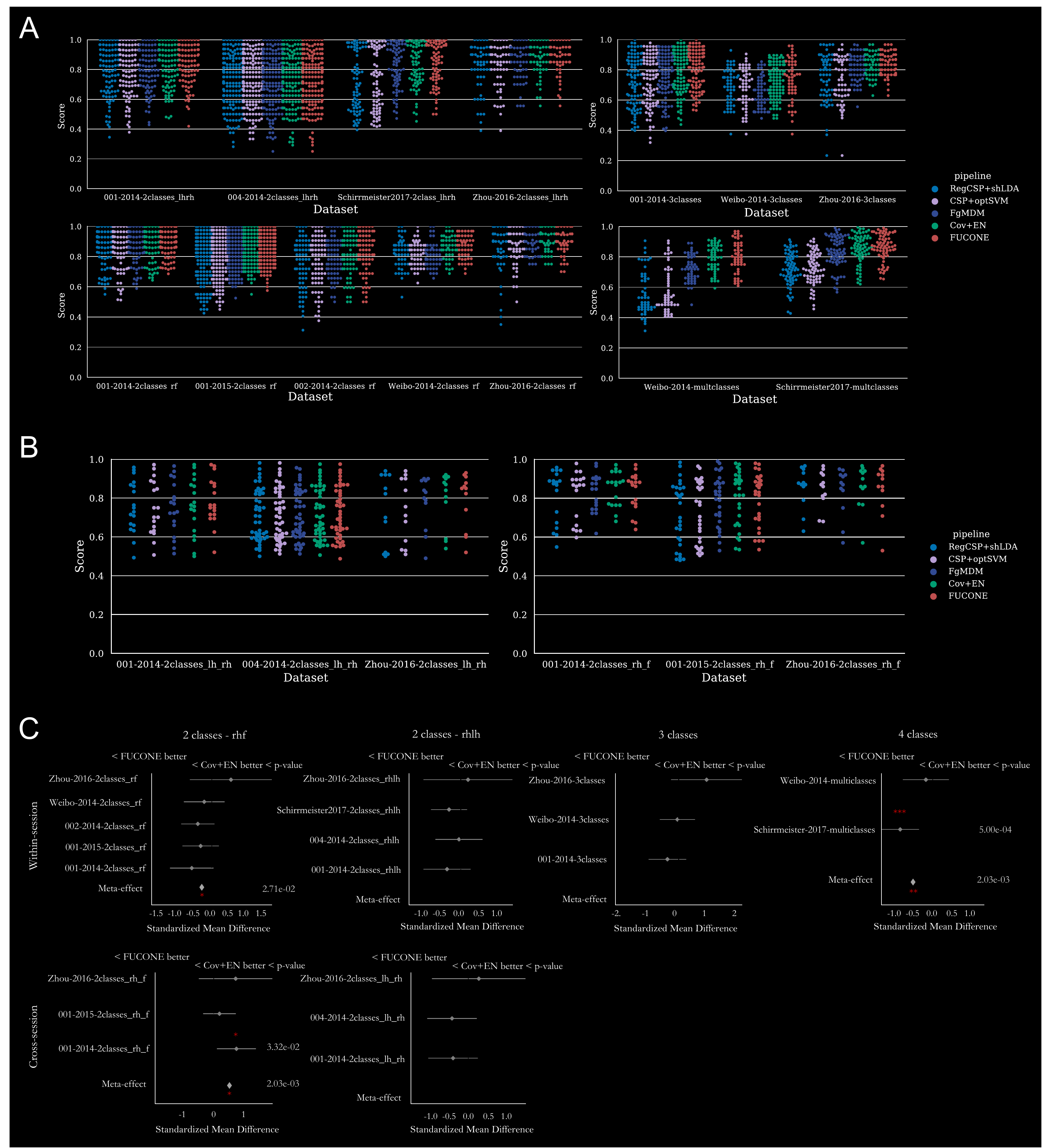}
		\caption{{Replicability assessments and comparison with state-of-the-art pipelines. (A) Within-session evaluation. On the left top, analysis performed with 2-class (lhrh stands for left hand vs right hand) datasets; on the left bottom, a comparison made with 2-class (rhf stands for right hand vs feet) datasets; on the right top, an analysis performed with 3-class datasets; on the right bottom, an analysis with 4-class datasets. (B) Cross-session evaluation. On the left, an analysis performed with 2-class (lhrh stands for left hand vs right hand) datasets; on the right, an analysis performed with 2-class (rhf stands for right hand vs feet) datasets. (C) Meta-effect assessment. Here we compared the two best pipelines: \cov{} and \fucone{}. Here, * stands for $p < 0.05$, ** for $p < 0.01$, and *** for $p < 0.001$.
		{\label{fig:meta}}%
		}}
\end{center}
\end{figure*}

From a more general perspective, given that the pipelines \cov{} and \fucone{} were the two best approaches, significantly better than the other state-of-the-art pipelines considered in this study (see Figures \textit{S2, S4, S6, S8, S10 \& S12}), and for a sake of brevity, we focused our statistical comparison to these two pipelines (see Figure~\ref{fig:meta}C). We observed that \fucone{} was significantly better than \cov{} in three cases: 2 classes "right hand vs feet" both in within-session ($p^{\text{meta}}=0.026$) and cross-session ($p^{\text{meta}}=0.023$), and 4 classes in within-session ($p^{\text{meta}}=0.002$). \fucone{} showed similar performance with \cov{} in the case of 2 classes "right hand vs left hand" (both in within and cross-session) and in the 3-class configuration (within session). For an exhaustive description of the statistical comparisons made between the whole set of pipelines, the reader can refer to Figures \textit{S1-12}.

\section{Discussion}
\subsection{Parameter optimization}
The proof-of-concept of \fucone{} have been tested during the IEEE WCCI Clinical BCI Competition~\cite{corsi_riemannian_2021}, achieving top results.
Still, several open questions were left in the aftermath of the competition. The first one was the potential improvement of the approach and complete evaluation of the parameters.
The second question was to investigate if the \fucone{} approach actually leads to an general improvement of the MI-based BCI performance or if its improvement is restricted only to the competition dataset. The last question dealt with the reasons behind this improved performance and the explanation of the higher scores.

To answer the first question, we listed the key parameters to be carefully tuned: the FC estimators, the frequency band of interest, and the pipeline combination to be taken into account by the ensemble method.

\textit{FC estimators} -- We preselected six complementary FC estimators (see Table \ref{tab:tab1}). By comparing the performance obtained at the first-level classification, we reduced our pool of estimators to the following metrics: the instantaneous coherence, the imaginary part of coherence, the phase-locking value and the debiased wPLI. Metrics derived from the coherence have previously proved to be of interest for MI-based BCI \cite{hamedi_electroencephalographic_2016, gonzalez-astudillo_network-based_2020} both in the classification \cite{mottaz_modulating_2018, cattai_phaseamplitude_2021} and in the study of mechanisms underlying the tasks performance \cite{corsi_functional_2020, corsi_bci_2021}. 
As refer to \plv, previous studies explored the possibility to use this estimator as alternative features with contradictory results. In, \cite{spiegler_phase_2004}, no particular changes were observed between mu rhythms in both hemispheres in phase coupling. In \cite{brunner_online_2006}, classification performance relying on power spectra features were better than those relying on \plv. In \cite{gysels_phase_2004}, \plv{} outperformed coherence and combining power with \plv{} features also gave the best performance. Finally, in \cite{yi_evaluation_2014}, the authors used \plv{} to identify patterns of simple and compound limb MI tasks. They notably elicited specific interactions between central and occipital areas within the $\theta$ band.
A reduced number of studies investigated the use of connectivity features derived from the phase-lag index. Recently in \cite{feng_functional_2020}, the authors compared the performance obtained from \pli, \wpli{} and \plv{} features. They observed that the FC features reached accuracy higher than 85\% and that \pli{} performed better than the filter-bank common spatial pattern.
In our study, we proposed to take into account complementary FC metrics in our ensemble method to take advantage of their physical properties. 

\textit{Frequency band} -- The second key parameter we optimized is the frequency band of interest. Surprisingly, any of the tested frequency bands led to significantly improved performance. An alternative approach would consist in considering a fine-tuned subject-specific system, in which each frequency band would be individually defined accordingly to the individual alpha peak \cite{klimesch_eeg_1999} as previously tested \cite{pichiorri_brain-computer_2015, corsi_functional_2020}. Even though this approach is particularly interesting and meaningful, it makes the interpretation harder at the meta-analysis level, especially in this work. To ensure the interpretability of our results, we preferred to work with broadband filters.

\textit{Ensemble} -- The last key parameter we tuned is the combination of pipelines to be taken into account by the ensemble method. The best configuration, corresponding to our \fucone{} approach, consisted of combining pipelines that respectively relied on instantaneous coherence, imaginary part of coherence, and covariance features. To our knowledge, there is no previous example of a combination of pipelines relying both on covariance matrices and connectivity estimations. Previous works investigated the possibility to combine FC with power spectra features, with contradictory outcomes~\cite{gysels_phase_2004, wang_phase_2006, krusienski_value_2012, cattai_phaseamplitude_2021}. Indeed, in~\cite{krusienski_value_2012}, pipelines relying on Fast Fourier Transform were not improved with PLV nor magnitude squared coherence, whereas in~\cite{wang_phase_2006}, the configuration combining large-scale synchrony and power features gave the best score. Here, to test the replicability of our results, we proposed to apply our \fucone{} approach a large number of datasets associated with MI tasks.

\subsection{Interpretation of discriminating connectivity features}

In the previous sections, we proved that our approach enabled a better discrimination between subjects' mental state, over a large number of datasets and configuration. Nevertheless, a remaining aspect to be taken into account is the variability, both in terms of BCI setup and at the individual level. 
Indeed, while having access to signals recorded by a large number of electrodes provides valuable information on brain connectivity, it engenders an increase of the experiment duration and of the subject's tiredness, as well as a decrease in Riemannian's performance~\cite{yamamoto_subspace_2021}. Defining a minimum number of electrodes to be used by our approach is necessary.

We proposed an additional step in the features selection consisting in a dimensional reduction (DR) approach. For a given training set, we performed a paired t-test to elicit the most discriminant interactions ($p < 0.05$ between the considered conditions). To ensure that the extracted features still consisted in symmetric positive definite matrices, we extracted the N channels that were the most frequently involved in discriminant interactions, meaning that they could be considered as hubs in terms of discriminant power. Finally, we took into account only the matrices (NxN) formed by these hubs. We tested our approach on a dataset for which an EEG montage of 128 electrodes was used \cite{schirrmeister_deep_2017} (for more details on the dataset, the reader can refer to Table~\ref{tab:tab2}). 

We tested different sets of channels ranging from 16 to 96 electrodes. First, we analyzed the performance at the group-level (see Figure \textit{S15}). Both for \imcoh{} and \instant{} the performance reached a plateau at $n_{\text{dr}}=32$, with an accuracy respectively of 0.675$\pm$0.123 and 0.754$\pm$0.14. These results indicated that only one fourth of the EEG montage set was sufficient to reach the maximum of the accuracy with \fucone{}. These methods gave access to the electrodes chosen over the subjects. We plotted the associated topographical maps that represented the occurrences over the subjects (see Figure \textit{S14}). 

We observed that at $n_{\text{dr}}=16$ the most frequently chosen channels were located above the bilateral sensorimotor area, which is in line with the areas involved during left- vs right-hand motor imagery. Then, as the number $n_{\text{dr}}$ increased, we observed a more diffused area selected by our approach, around the core one that consisted in the regions associated with the configuration $n_{\text{dr=16}}$. These more diffused regions were located above the parietal and occipital areas, both related to associative areas, known to play a crucial role in motor sequence learning as well as in abstract task learning~\cite{mcdougle_taking_2016,hetu_neural_2013,hardwick_neural_2018,dayan_neuroplasticity_2011} and in MI tasks performance~\cite{guillot_neurophysiological_2010}. 

These observations are in line with previous works conducted with functional connectivity to elicit markers of motor imagery-based BCI performance~\cite{corsi_functional_2020}.
At the individual level, we observed the "plateau effect" for ten to fourteen subjects (see Figure~\textit{S16}). For the remaining four subjects, two cases were observed: an increase of the performance with a larger number of electrodes considered (subjects 3, 4 and 8) or a decrease of the performance above $n_{\text{dr}}=32$ (subject 1). This last point means that considering the DR approach can be valuable to individually adapt the number of considered electrodes.

\subsection{Study on post-stroke patient}

Another crucial element is the inter- and intra-subject variability. This is particularly true for clinical applications where attention must be paid to customize the BCI system to the patients' ability to perform a motor imagery task and their neurophysiological signature~\cite{scherer_individually_2015} (\emph{e.g.}, brain lesions caused by the stroke can engender a modification of the spatial and frequency locations of the features).

Our DR approach can provide a valuable tool to take into account the patient's specificity. We tested our pipeline with one of the patients who were the least responsive to the BCI training in~\cite{scherer_individually_2015}: patient A (42 years old) who suffered from a locked-in syndrome due to brainstem stroke. Among the different motor imagery tasks proposed, the poorer results were obtained when he performed the right hand vs feet motor imagery. We applied our approach on the EEG recordings and compared our results to the state-of-the-art pipelines (see Figure \textit{S17A}). We slightly modified the pipeline described in the previous paragraph so that the algorithm chose the most suited number of electrodes (ranging from 10\% to 30\% of the available set of electrodes) according to the performance obtained on the training set. 

Interestingly, our approach based on the combination of \fucone{} with DR showed the best results (0.54$\pm$0.09) and performed better than \fucone{} taken alone (0.45$\pm$0.17). This last point highlights the need to take into account the patient's specificity in the classification pipeline. More importantly, as previously explained, DR gives access to the electrodes that are selected for the classification (see Figure \textit{S17B}). In our example, we observed that among the 30 EEG electrodes, the most suited electrodes for \imcoh{} were CP3 and P4 and that with \instant, we selected C4 and O1. The location of the preselected channels, mostly above the sensorimotor area and the associative areas, reinforces the idea that our approach enables a relevant selection of channels, in line with previous works based on functional connectivity~\cite{yi_evaluation_2014, corsi_functional_2020}.

Nevertheless, we are perfectly aware that applying the DR approach after \fucone{} consists in an additional computation step. Therefore, we think that it should be considered in specific cases: to identify the most suited electrodes to be used and for patients who are less responsive to motor imagery-based BCI.

\subsection{Characterizing potential contributions of functional connectivity} 

One important improvement of \fucone{} is to enhance the results for the most difficult cases, that is for the least responsive subjects. Those subjects obtain very low classification accuracy with all state-of-the-art methods and are the focus of many research works~\cite{jeunet_predicting_2015,corsi_bci_2021}. Our approach systematically increase the average accuracy for these subjects, for eight datasets and different MI classes, as shown on Figure \textit{S13}. The reason behind this improvement are mostly imputable to the robustness brought by the combination of multiple feature spaces (covariance and FC) to yield the final prediction.

To reveal insights on the situations in which FC estimators can capture additional information with respect to the covariance matrices, we worked with a toy model. For this purpose, we simulated EEG signals via the following process. First, we generated two sinus signals, referred as sources. To obtain four synthetic EEG signals, we applied a mixing matrix to these sources, and we added some noise to each signal (see Figure \textit{S18}A). From these synthetic EEG signals, we computed respectively the associated covariance, coherence, and imaginary part of coherence matrices (see Figure \textit{S18}A). The color scheme indicates the matrix entry values and it can be seen that Cov, Coh and ImCoh capture different relationship between the synthetic EEG signals. They are considered as references and denoted $E_{\mathrm{ref}}$. The next step consisted of studying the effect of three parameters: the phase, the amplitude, and the frequency. To this end, we varied each of these parameters separately by modifying the properties of one of the two sources. To assess to which extent an estimator E is sensitive to the variation of a given parameter p, we computed the Riemannian distance between $E$ and $E_{ref}$.
First, we observed that the coherence (\instant{}) and the imaginary part of coherence (\imcoh{}) are complementary (see Figure \textit{S18}B).  Indeed, \imcoh{} is sensitive to changes in phase and not in amplitude whereas \instant{} shows the opposite trend. These results are in line with previous findings showing the complementary behavior of the coherence and the imaginary part of the coherence \cite{cattai_phaseamplitude_2021}. As refer to the covariance (\cov{}), it seems that it is more sensitive towards changes in phase and in amplitude than the FC estimators. Nevertheless, in the case of source that varies in frequency, \instant{} presents a higher variation than \cov{}. \instant{} shows a sensitivity to frequency variation, while \cov{} and \imcoh{} are mostly invariant to change in frequency (see Figure \textit{S18}B). Therefore, \cov{}, \instant{} and \imcoh{} capture each one a different aspect of the signal, being sensitive or invariant to variation in phase, amplitude, and frequency. This diversity could help to recover discriminative information when an oscillatory activity is involved, which is precisely the case when performing a motor imagery task. Therefore, FC estimators could recover more discriminative information than \cov{} when an oscillatory activity is involved, which is precisely the case when performing a motor imagery task.

To further investigate the FC estimator contribution to covariance-based pipelines, we also tested the diversity and reported the results on BNCI001-2015 in Supplementary material (see Figures \textit{S19-20}). We also tested other datasets and the results were significantly the same across datasets.
The instantaneous coherence had a very low relative diversity and potential improvement when compared to \imcoh, \plv, \pli, \wpli{} and \aec{} estimators. The \imcoh{} estimator showed the highest relative diversity and potential improvement, but \plv, \pli, \wpli{} and \aec{} were very close in absolute value.

It is interesting to see that the instantaneous coherence displayed the lowest diversity score whereas it was the FC estimator achieving the highest accuracy, after \cov{} and that other FC estimators showed high diversity and low classification accuracy. The diversity score reflects the potential improvement that is possible to achieve by combining several estimators. As \fucone{} relies on an ensemble classifier that evaluates the confidence of each pipeline, FC-based pipelines with low accuracy have a low confidence even if their diversity scores are high. This point out that there is a potential room to improve the combination of FC estimators and covariance measure. A feature space that builds upon the careful aggregation of those estimators could have a real impact on classification performance. The most difficult part of this challenge is either to find the correct geometry to define such feature space or to be able to assess the reliability of the FC pipelines.

\subsection{Caveats and limitations}
Even though our approach enabled an improvement of the BCI accuracy over a large number of datasets, this study presents several caveats that need to be acknowledged. 

A first limitation is related with the requirements made for estimating covariance or FC features. The estimators need to provide symmetric and positive definite matrices in order to apply the Riemannian framework. We ensured this critical point by projecting onto the manifold the matrices that were ill-conditionned, \emph{i.e.}, with eigenvalues closed to zero~\cite{chevallier_riemannian_2018}. This is a common problem for covariance estimators when a Common Average Reference (CAR) is applied or when using a combination of working electrodes to estimate the signal of a disconnected electrode. We did not use any electrode re-referencing or missing signal completion in this work, and this was sufficient to ensure that our approach was working properly. For low quality EEG acquisition with missing data, it is still possible to rely on robust estimation to obtain accurate results~\cite{yger_geodesically-convex_2020}.

Secondly, we still observed an inter-subject variability in terms of classification performance. Several elements can explain it.
In an effort to test our approach in conditions closest to real-life scenarios, we applied the \fucone{} method on datasets that 
presented a strong diversity in terms of MI tasks and of number of channels considered without preprocessing. 
The inter-subject variability can reflect a variety in the possible ways to detect neurophysiological properties underlying the MI performance. We led our analysis by considering both covariance and FC features. However, further analysis, possibly in the source space, could be of interest to provide a more accurate description of the neural mechanisms~\cite{barzegaran_functional_2017}, notably those underlying the control of a BCI~\cite{corsi_functional_2020}. Our \fucone{} approach being new, the aim of this study was to assess to which extent our approach could be robust enough towards a variability in terms of datasets, i.e.  in terms of experimental setup. A further study will consist of exploring new ways to compute functional connectivity estimators. We plan in a next future to apply our approach in the source space, on a dedicated dataset.

This inter-subject variability is strongly linked to the BCI inefficiency phenomenon that is a multifaceted problem that cannot be only addressed by improving our way to discriminate the subjects' mental state. Indeed, machine-centered and user-centered approaches, relying on the subjects' similarities~\cite{khazem_minimizing_2021} or on the subjects' cognitive profile~\cite{jeunet_predicting_2015} for example, should be considered together to improve BCI systems~\cite{thompson_critiquing_2018}. 

Lastly, the \fucone{} approach has only been tested offline. To propose a reliable alternative to current classification pipelines, an online implementation is required. Before considering testing it online, we wanted first to ensure that the \fucone{} method was replicable. Nevertheless, we led a feasibility study and we observed that our approach may engender a delay caused by the processing time. Such a delay strongly depends on the time window of interest used to extract the FC features. In this work, we chose to work with epochs of 1s. This led to a processing time in the order of 100 ms, that is compatible for an online implementation. By working with overlapping epochs, we can reduce the time delay in order to refresh the feedback more often and ensure the subject's sense of agency following a similar approach to~\cite{kalunga_online_2016}.

\section{Conclusion}

In this work, we introduced our approach, named \fucone{}, that consists of a Riemannian take on functional connectivity estimators, by combining the decision in each feature space with an ensemble classifier. We thoroughly optimized the key parameters on a reference dataset. We evaluated \fucone{} on eight MI datasets, considering different imaginary movement combinations and evaluating results when trained on the same session (within-session evaluation) or trained on different sessions (cross-session evaluation). Consistent with our hypothesis, our approach was significantly better than several state-of-the-art methods, either CSP-based or covariance-based Riemannian classifiers.
An extension of our method, relying on a dimensional reduction, elicited a subset of electrodes, located above sensorimotor and associative areas, that were sufficient to achieve high accuracy. Such a technique proved to be relevant for clinical applications. A characterization of the contribution of functional connectivity estimators to covariance-based pipeline led us to point out that there is room for a potential improvement of the combination of decision. Taken together, our results offer new insights into the need to consider functional connectivity based methods to improve the BCI performance.

\section{Acknowledgments}
FDVF and MCC acknowledge support from the European Research Council (ERC) under the European Union’s Horizon 2020 research and innovation programme (grant agreement No. 864729). Their research leading to these results has received funding from the program "Investissements d’avenir" ANR-10-IAIHU-06.
FY acknowledges the support of the ANR as part of the ``Investissements d'avenir'' program, reference ANR-19-P3IA-0001 (PRAIRIE 3IA Institute).
The funders had no role in study design, data collection and analysis, decision to publish, or preparation of the manuscript.

\section{Data and code availability}
The code used to perform the analysis and the data that support the findings of this study are publicly available in this Github repository: \url{https://github.com/mccorsi/FUCONE.git} (accessed on December 17 2021).

\section{Authors contributions}
MCC, SC, and FY initiated research; MCC, SC, FY and FDVF designed research; MCC, SC, and FY performed research; MCC, SC, and FY contributed analytic tools; MCC and SC analyzed data; and MCC, SC, FY and FDVF wrote the paper. All authors revised and approved the manuscript.

\section{Additional Information}
Supplementary Information accompanies this paper.

\bibliographystyle{nature_mag}
\bibliography{fucone_paper.bib}

\selectlanguage{english}
\clearpage
\end{document}


\captionsetup[figure]{name=Figure}
\captionsetup[table]{name=Table}
\renewcommand\thefigure{S\arabic{figure}}
\renewcommand\thetable{S\arabic{table}}

\newcommand{\suppression}[1]{\textcolor{blue}{\st{#1}}}


\title{Functional connectivity ensemble method to enhance BCI performance (FUCONE)\footnote{This work has been submitted to the IEEE for possible publication. Copyright may be transferred without notice, after which this version may no longer be accessible.}}
\author{M.-C. Corsi, S. Chevallier, F. De Vico Fallani and F. Yger}

\maketitle

\section*{Supplementary Figures}
 
\begin{figure}[ht!]
\begin{center}
\includegraphics[width=\textwidth]{figures/Supplem/SI_1_revised.pdf}
\caption{Meta-analysis performed over four datasets, 2 classes (left- vs. right-hand motor imagery) with a within-session evaluation. Here, * stands for $p<0.05$, ** for $p<0.01$, and *** for $p<0.001$.}
\label{Figure1}
\end{center}
\end{figure}

\newpage
\begin{figure}[ht!]
\begin{center}
\includegraphics[width=\textwidth]{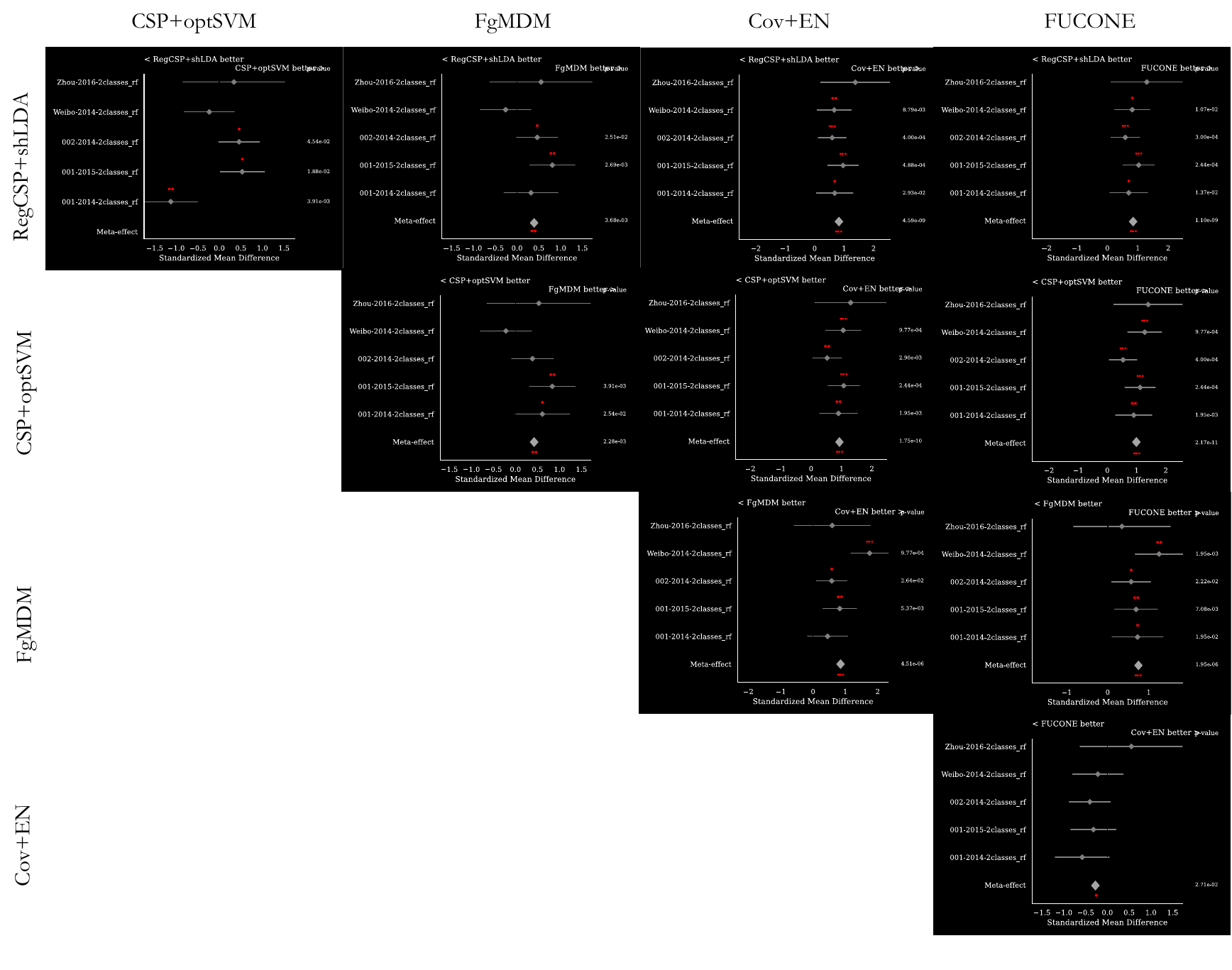}
\caption{Meta-analysis performed over four datasets, 2 classes (left- vs. right-hand motor imagery) with a within-session evaluation. Here, * stands for $p<0.05$, ** for $p<0.01$, and *** for $p<0.001$.}
\label{Figure2}
\end{center}
\end{figure}
\newpage

\begin{figure}
\centering
\includegraphics[width=\textwidth]{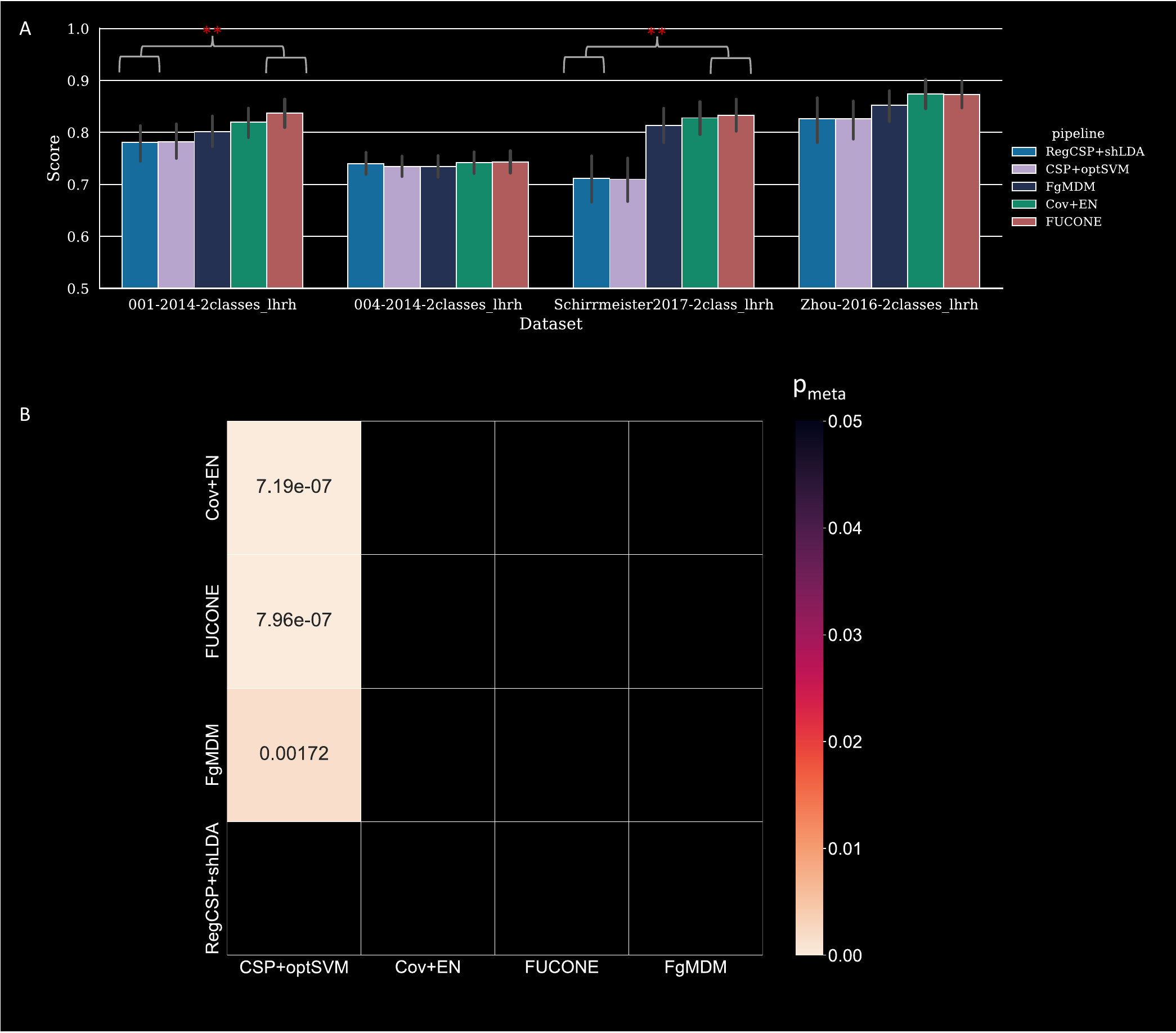}
\caption{Meta-analysis performed over five datasets, 2 classes (feet vs. right-hand motor imagery) with a within-session evaluation. Here, * stands for $p<0.05$, ** for $p<0.01$, and *** for $p<0.001$.}
\label{Figure3}
\end{figure}
\newpage

\begin{figure}
\centering
\includegraphics[width=\textwidth]{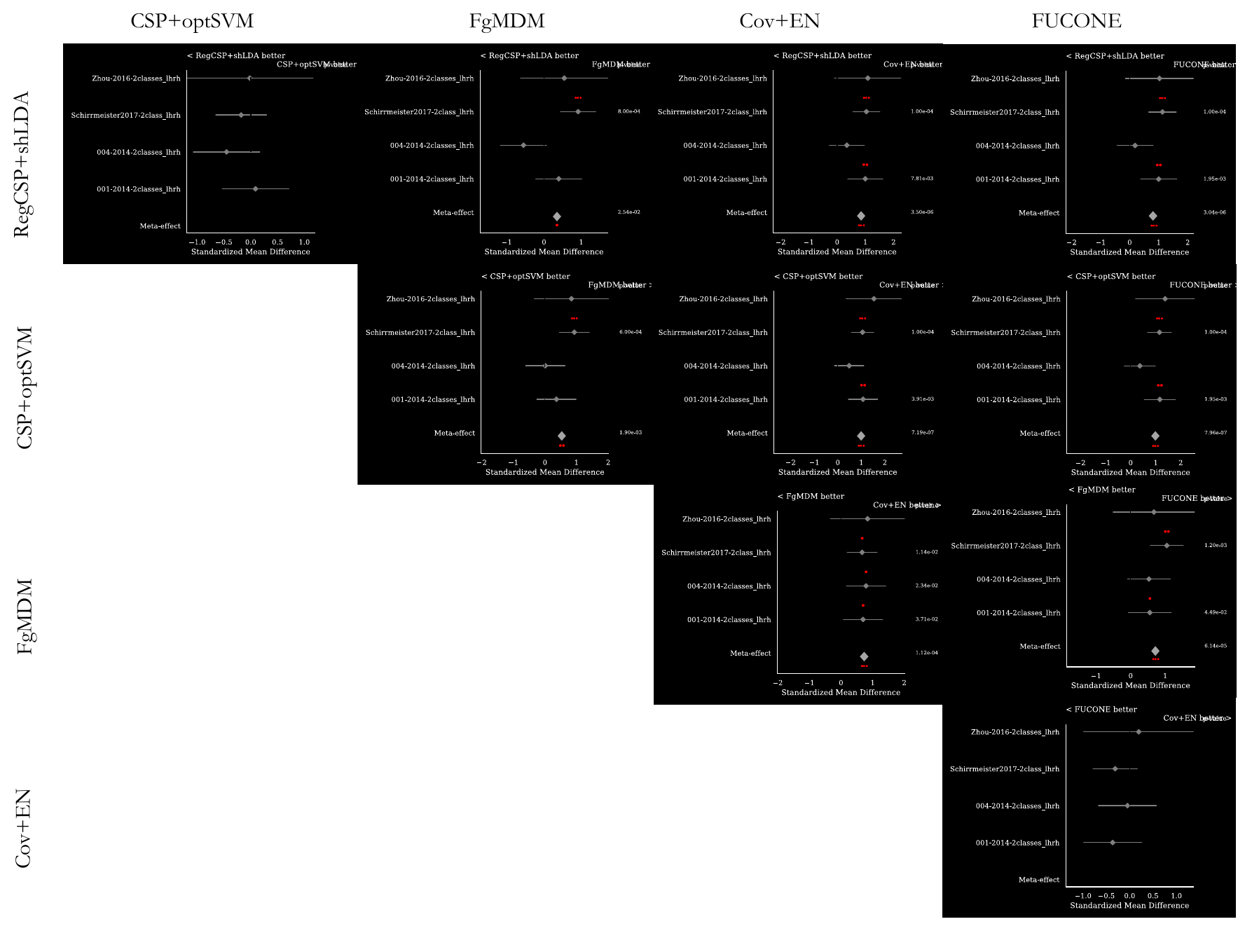}
\caption{Meta-analysis performed over five datasets, 2 classes (feet vs. right-hand motor imagery) with a within-session evaluation. Here, * stands for $p<0.05$, ** for $p<0.01$, and *** for $p<0.001$.}
\label{Figure4}
\end{figure}
\newpage

\begin{figure}
\centering
\includegraphics[width=\textwidth]{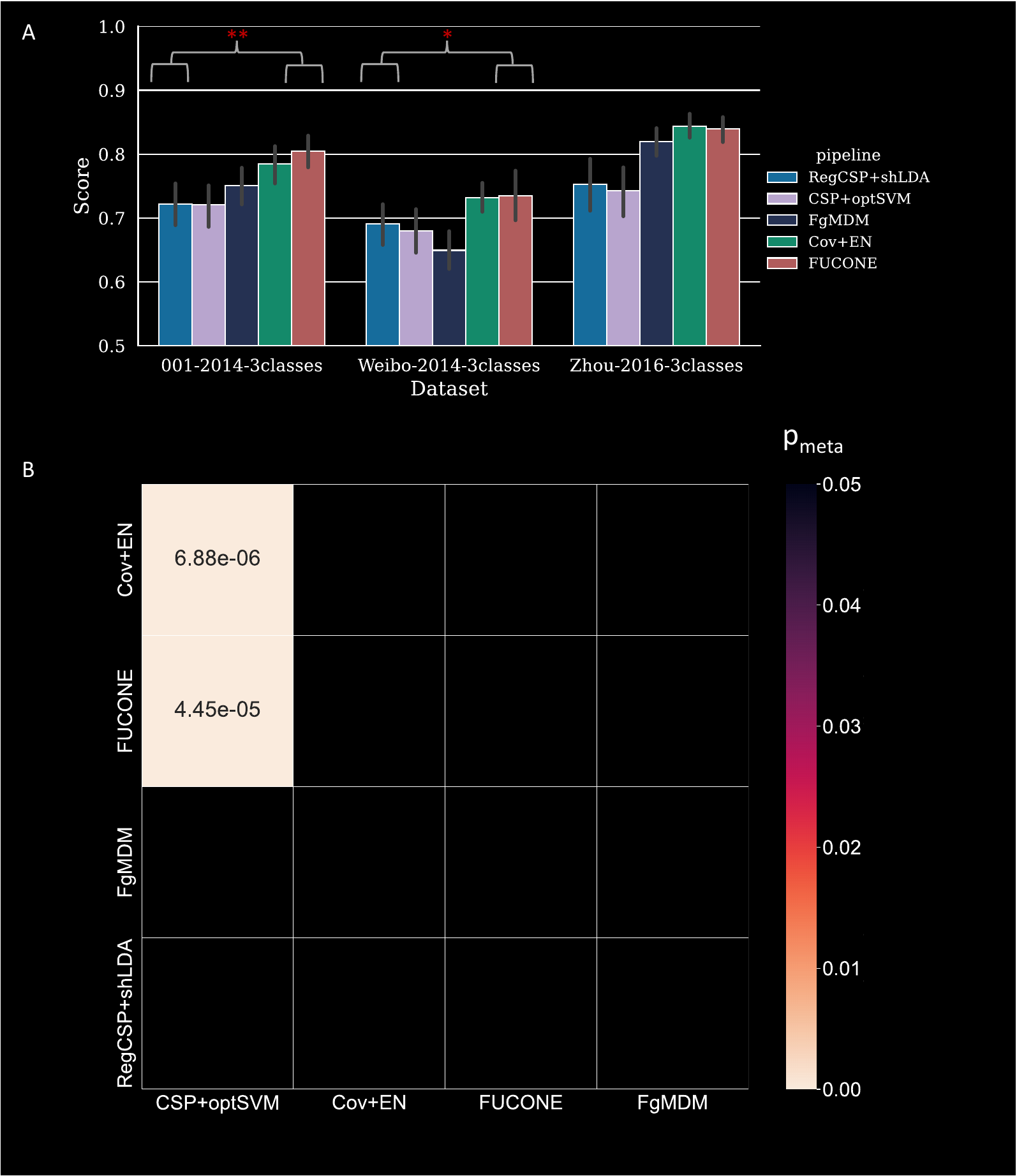}
\caption{Meta-analysis performed over three datasets, 3 classes with a within-session evaluation. Here, * stands for $p<0.05$, ** for $p<0.01$, and *** for $p<0.001$.}
\label{Figure5}
\end{figure}
\newpage

\begin{figure}
\centering
\includegraphics[width=\textwidth]{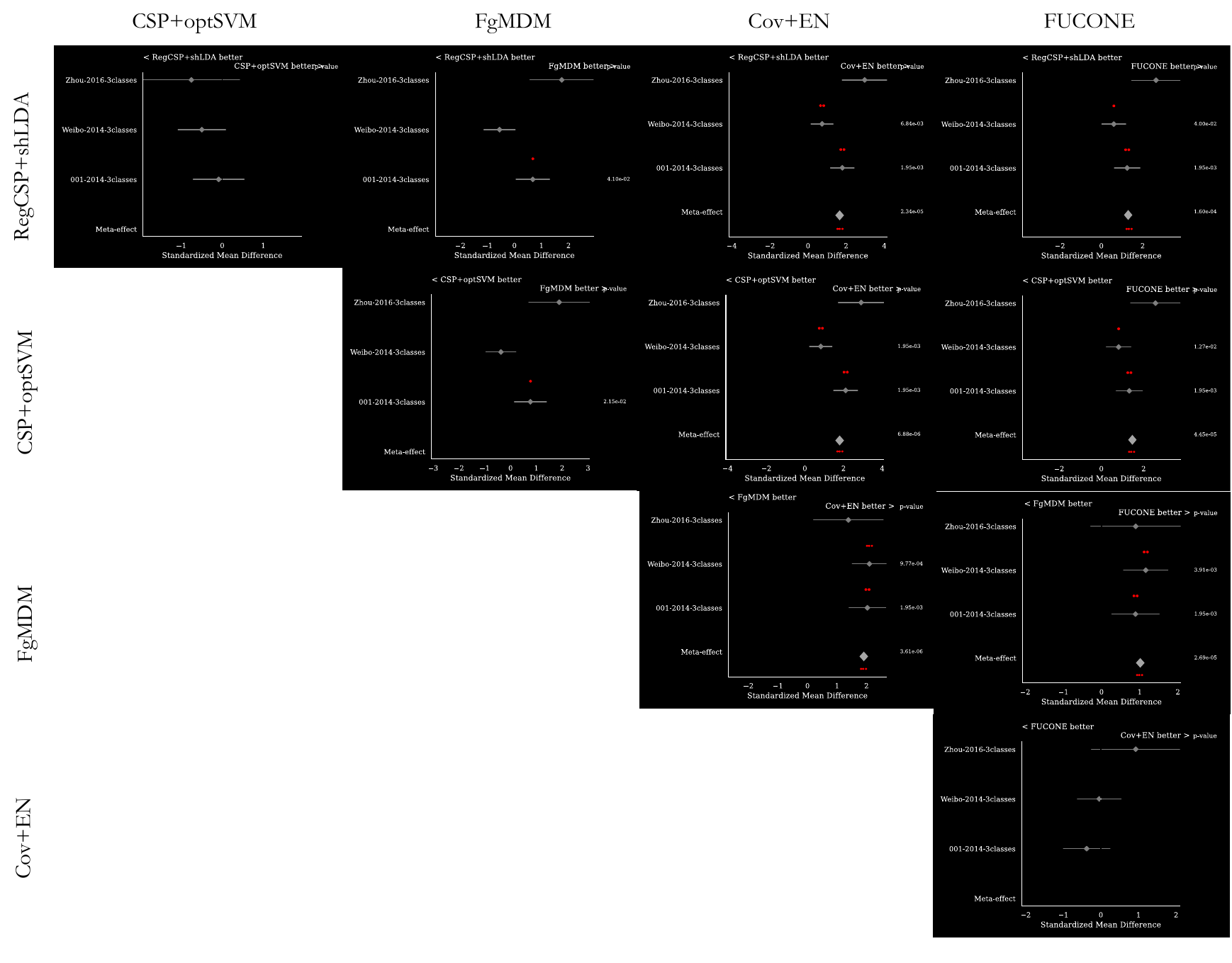}
\caption{Meta-analysis performed over three datasets, 3 classes with a within-session evaluation. Here, * stands for $p<0.05$, ** for $p<0.01$, and *** for $p<0.001$.}
\label{Figure6}
\end{figure}
\newpage

\begin{figure}
\centering
\includegraphics[width=\textwidth]{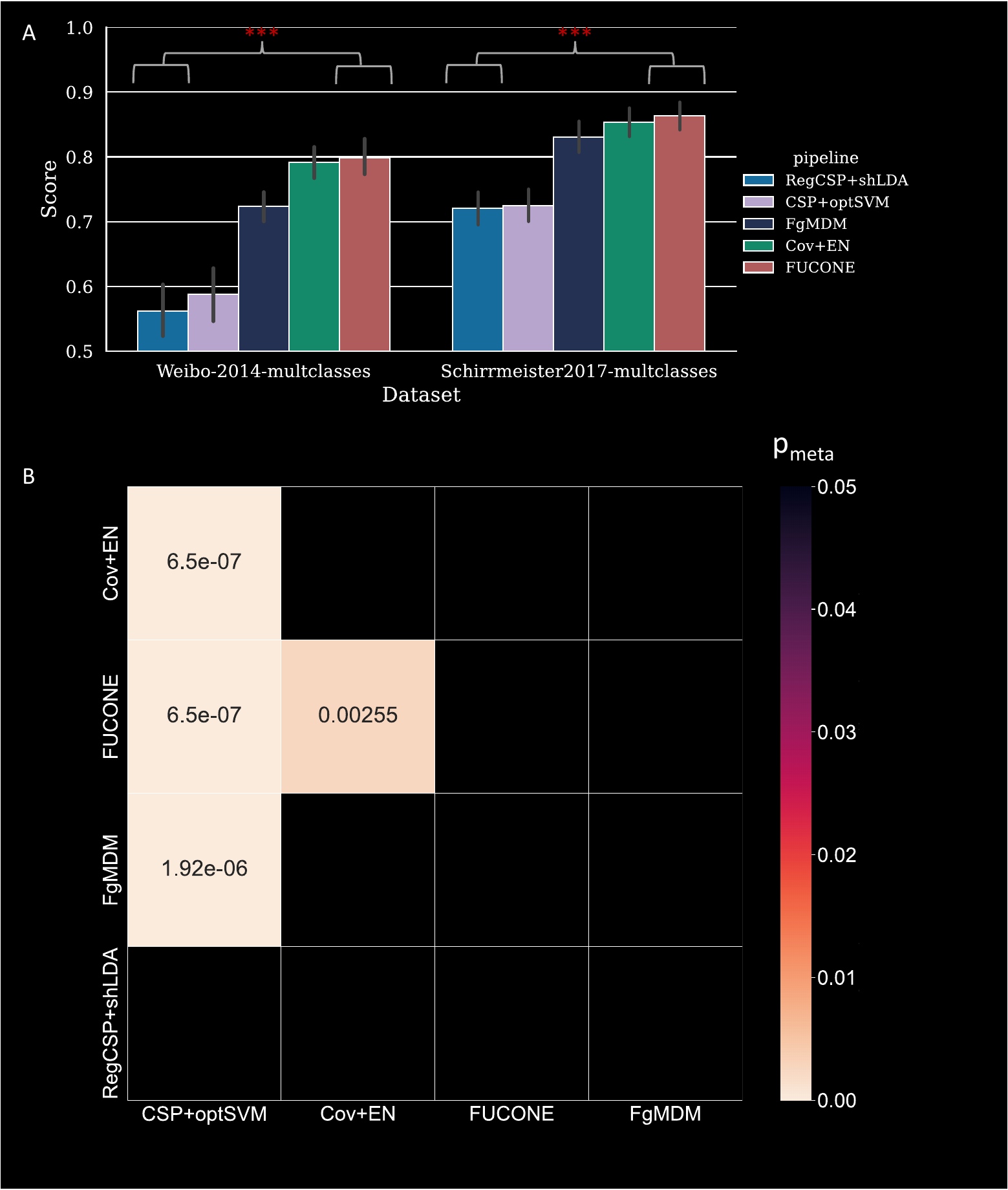}
\caption{Meta-analysis performed over two datasets, 4 classes with a within-session evaluation. Here, * stands for $p<0.05$, ** for $p<0.01$, and *** for $p<0.001$.}
\label{Figure7}
\end{figure}
\newpage

\begin{figure}
\centering
\includegraphics[width=\textwidth]{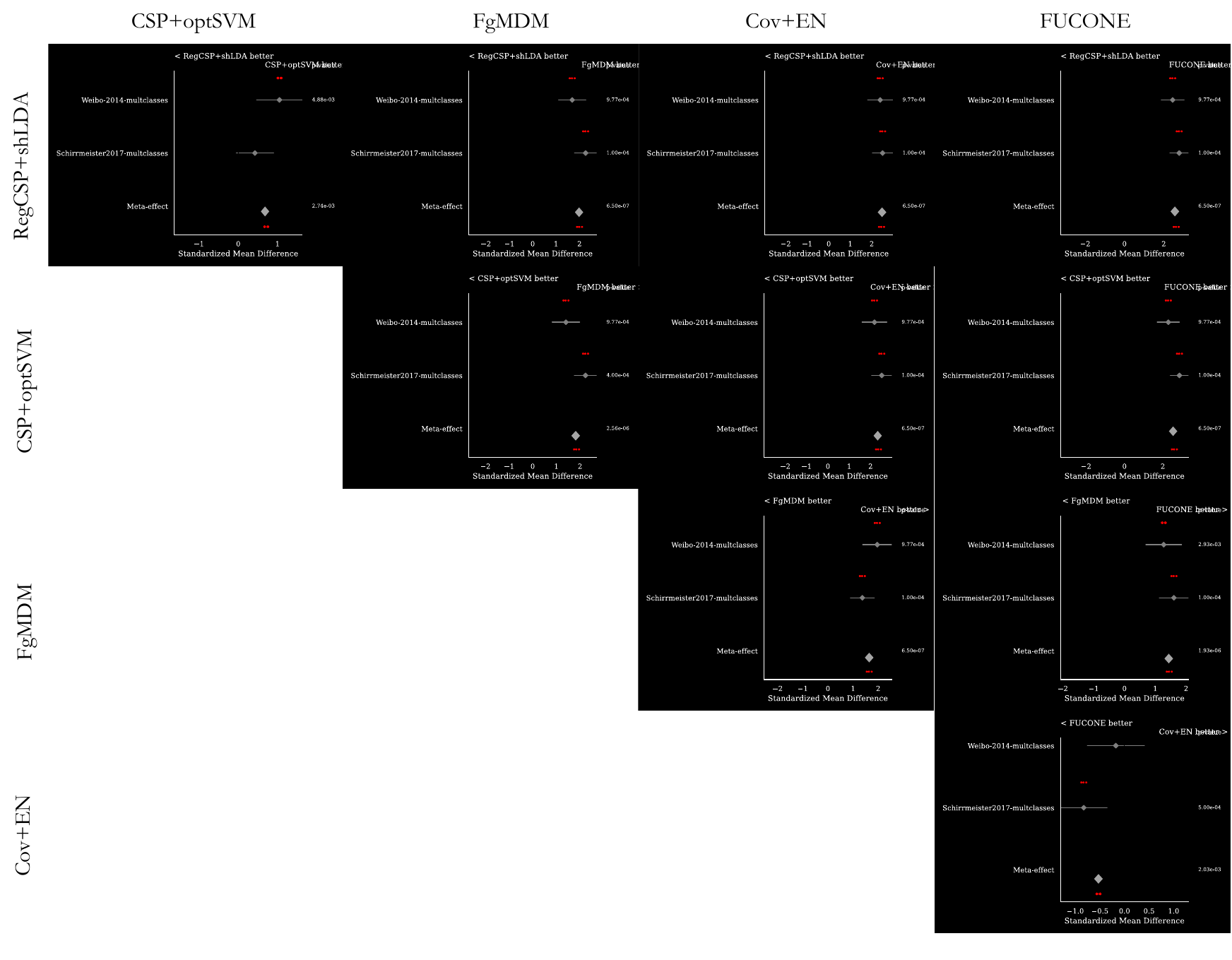}
\caption{Meta-analysis performed over two datasets, 4 classes with a within-session evaluation.}
\label{Figure8}
\end{figure}
\newpage

\begin{figure}
\centering
\includegraphics[width=\textwidth]{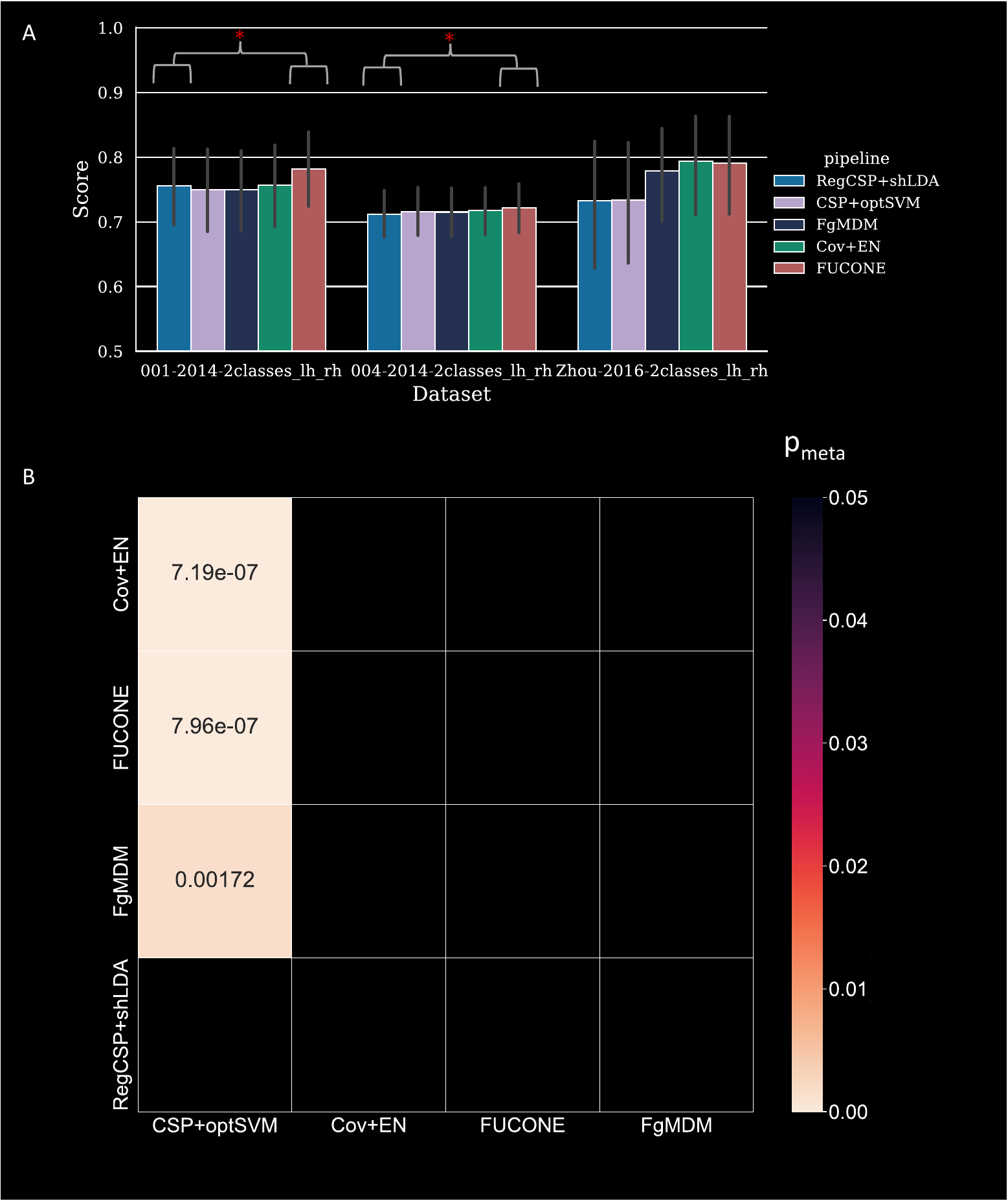}
\caption{Meta-analysis performed over three datasets, 2 classes (left- vs. right-hand motor imagery) with a cross-session evaluation. Here, * stands for $p<0.05$, ** for $p<0.01$, and *** for $p<0.001$.}
\label{Figure9}
\end{figure}
\newpage

\begin{figure}
\centering
\includegraphics[width=\textwidth]{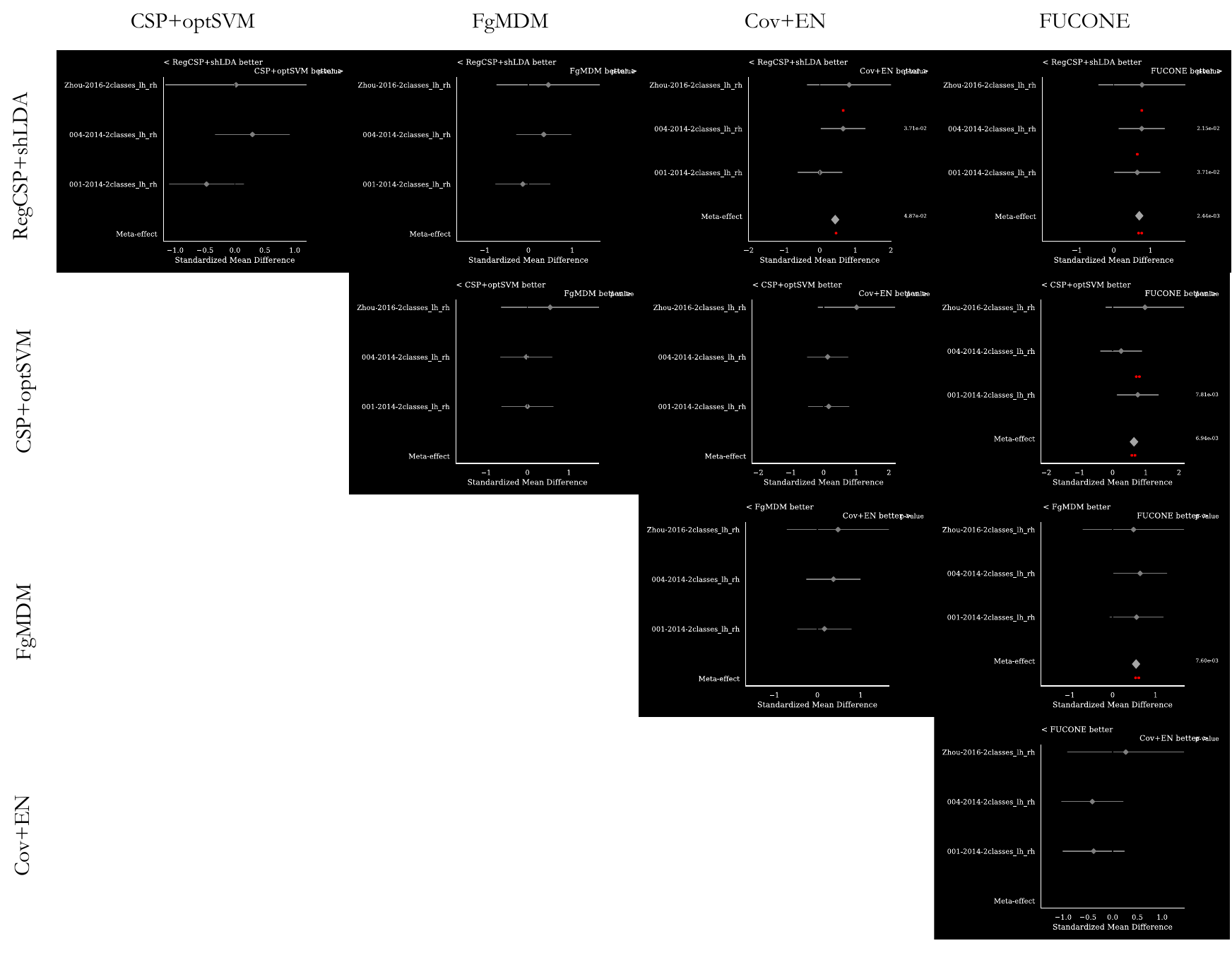}
\caption{Meta-analysis performed over three datasets, 2 classes (left- vs. right-hand motor imagery) with a cross-session evaluation. Here, * stands for $p<0.05$, ** for $p<0.01$, and *** for $p<0.001$.}
\label{Figure10}
\end{figure}
\newpage

\begin{figure}
\centering
\includegraphics[width=\textwidth]{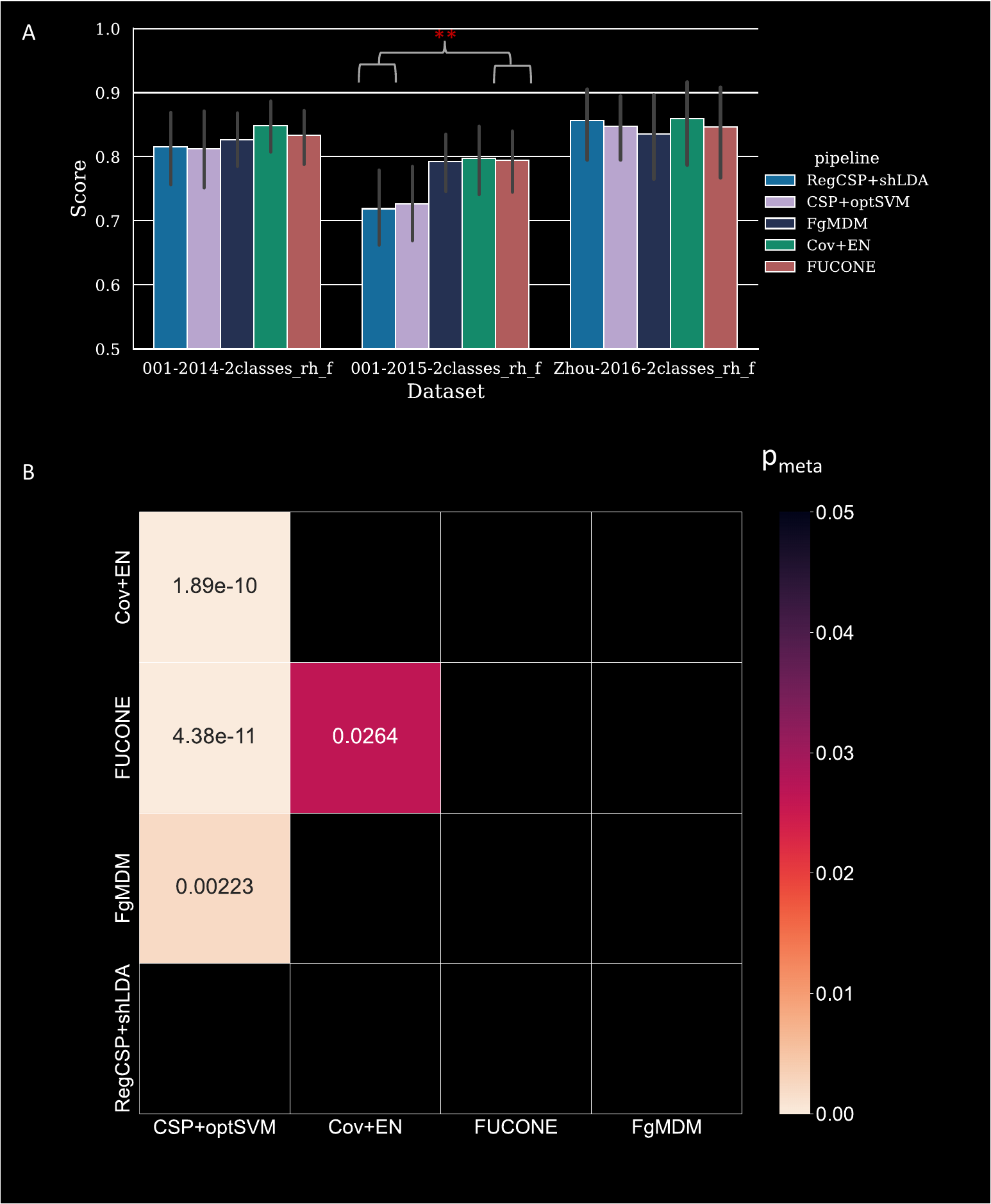}
\caption{Meta-analysis performed over three datasets, 2 classes (feet vs right-hand motor imagery) with a cross-session evaluation. Here, * stands for $p<0.05$, ** for $p<0.01$, and *** for $p<0.001$.}
\label{Figure11}
\end{figure}
\newpage

\begin{figure}
\centering
\includegraphics[width=\textwidth]{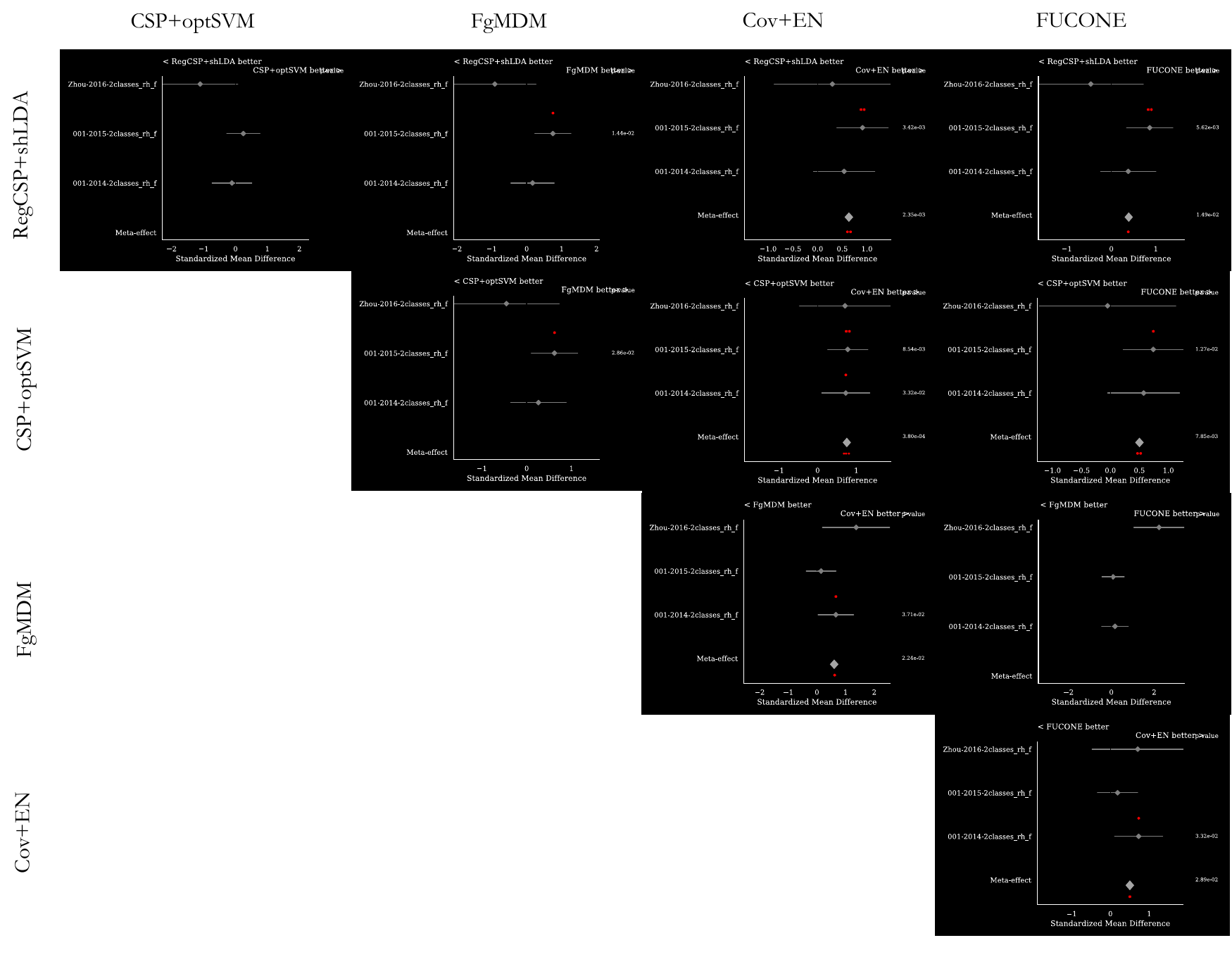}
\caption{Meta-analysis performed over three datasets, 2 classes (feet vs right-hand motor imagery) with a cross-session evaluation. Here, * stands for $p<0.05$, ** for $p<0.01$, and *** for $p<0.001$.}
\label{Figure12}
\end{figure}
\newpage

\begin{figure}
\centering
\includegraphics[width=\textwidth]{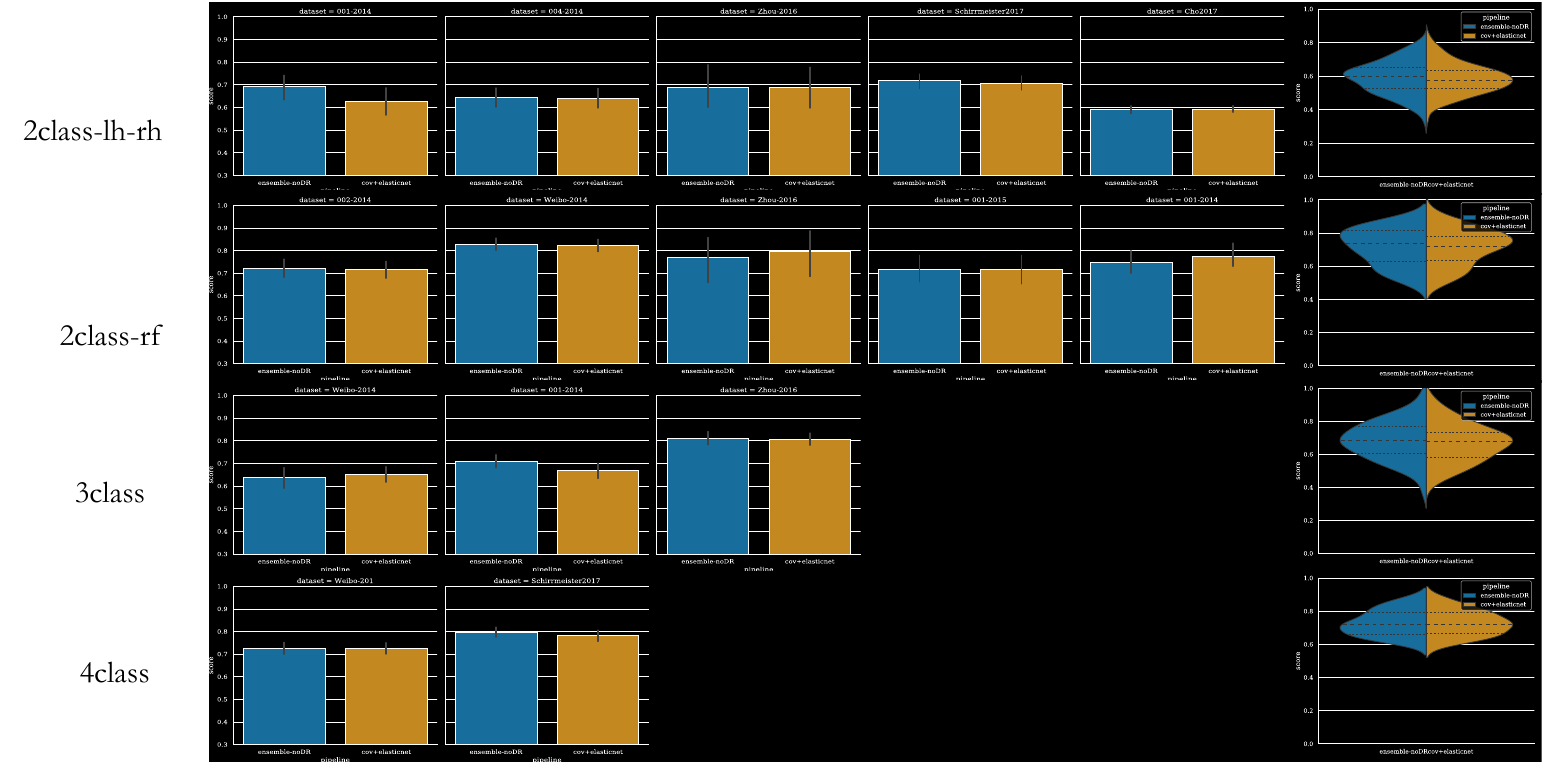}
\caption{Comparison of FUCONE with Cov+EN in the least responsive subjects (\emph{i.e.}, 30\% subjects who presented the weakest performance, both at the single dataset level and at a global over for a given configuration (\emph{i.e.}, number of tasks). }
\label{Figure13}
\end{figure}
\clearpage
\newpage

\begin{figure}
\centering
\includegraphics[width=\textwidth]{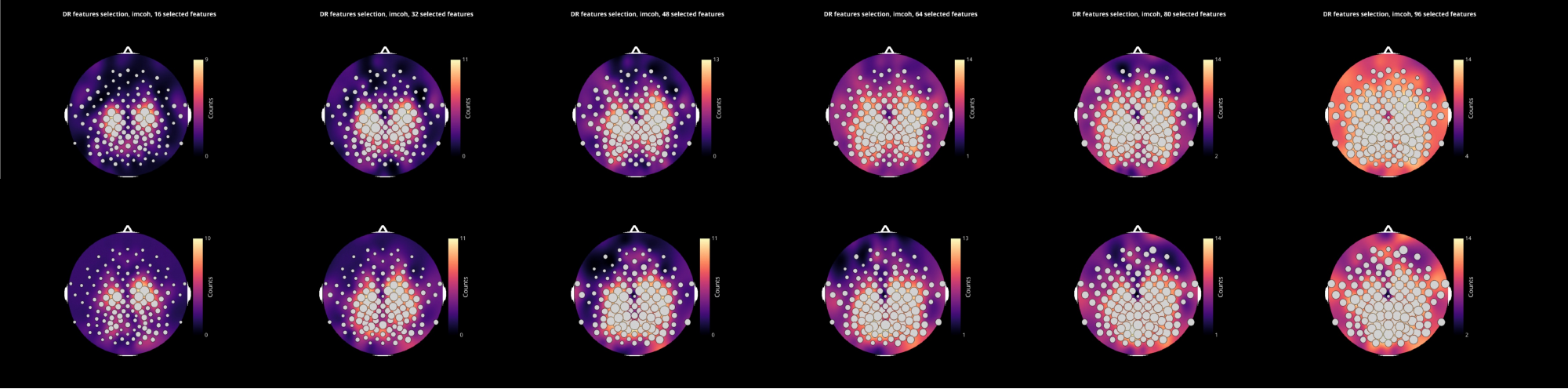}
\caption{Evolution of the features selection of an increasing number of channels considered. Each topography represents the occurrence associated to each sensor over the fourteen subjects included in the dataset from Schirrmeister et al.
}\label{Figure14}
\end{figure}
\newpage

\begin{figure}
\centering
\includegraphics[width=\textwidth]{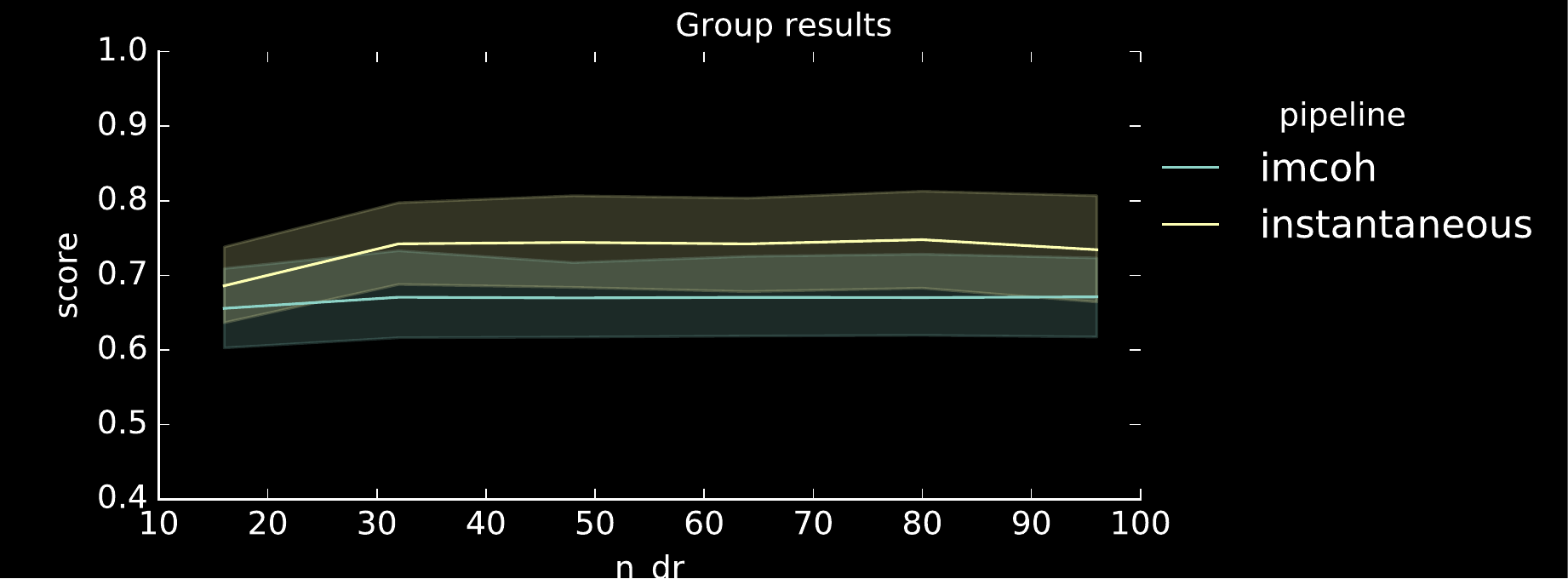}
\caption{Evolution of the performance with an increasing number of channels considered. It corresponds to the group analysis performed over the fourteen subjects included in the dataset from Schirrmeister et al.
}\label{Figure15}
\end{figure}

\begin{figure}
\centering
\includegraphics[width=\textwidth]{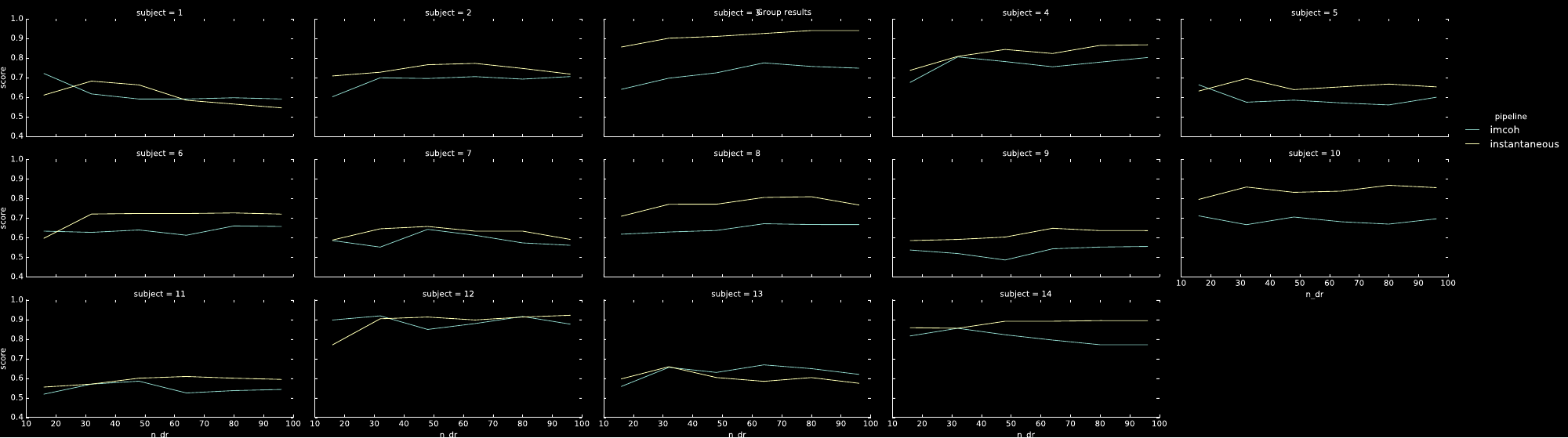}
\caption{Evolution of the performance with an increasing number of channels considered. It corresponds to the individual analysis performed over the fourteen subjects included in the dataset from Schirrmeister et al.
}\label{Figure16}
\end{figure}

\begin{figure*}
\centering
		\includegraphics[width=450pt]{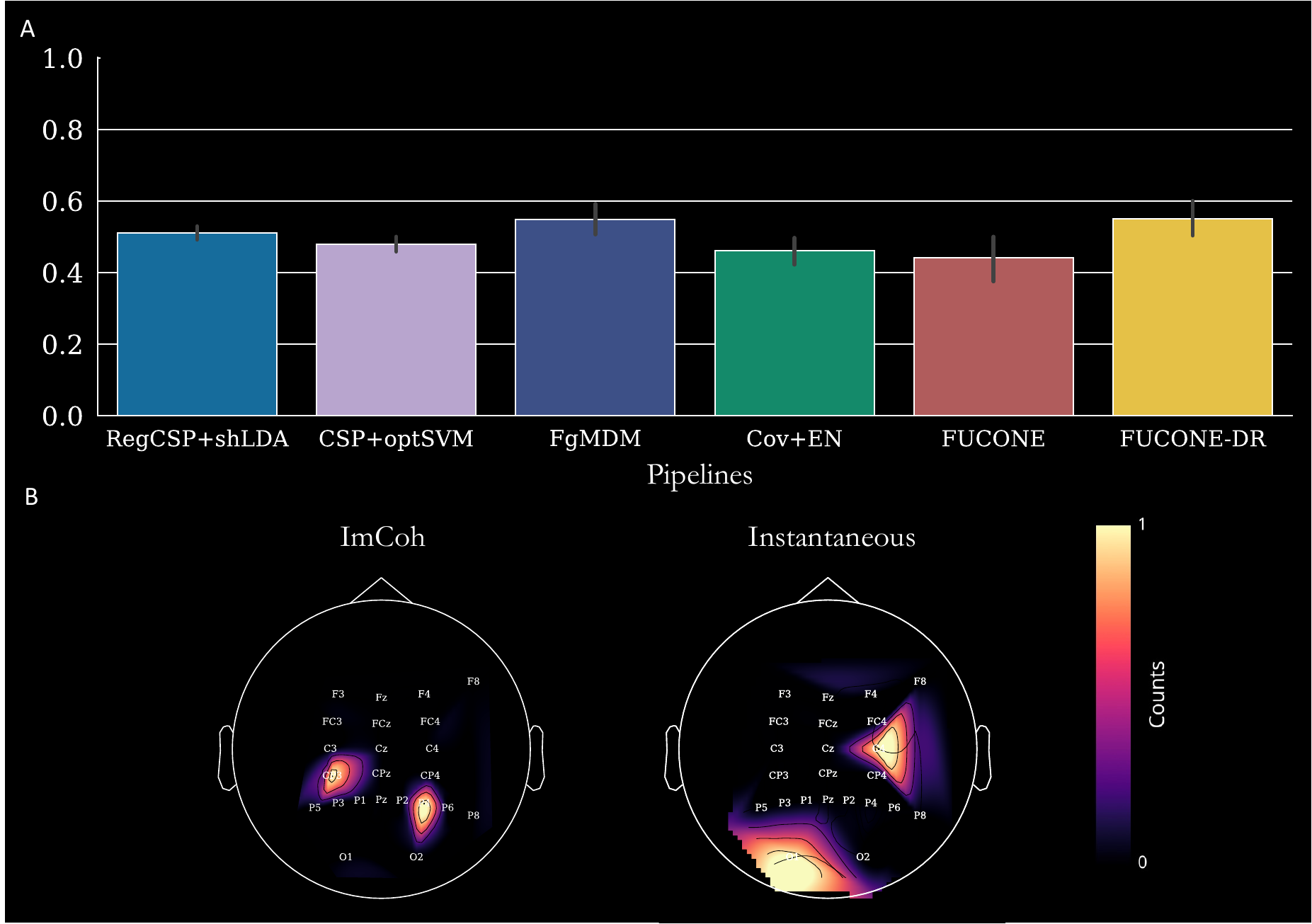}
		\caption{Clinical application of our approach. (A) Comparison of the performance obtained from our FUCONE approaches and the state-of-the-art pipelines. We used the dataset provided by Scherer et al. More particularly, we chose the patient who was the least responsive in  right hand vs feet motor imagery tasks. (B) Electrodes selected by our DR approach.
		{\label{Figure17}}%
		}
\end{figure*}

\begin{figure*}
\centering
		\includegraphics[width=450pt]{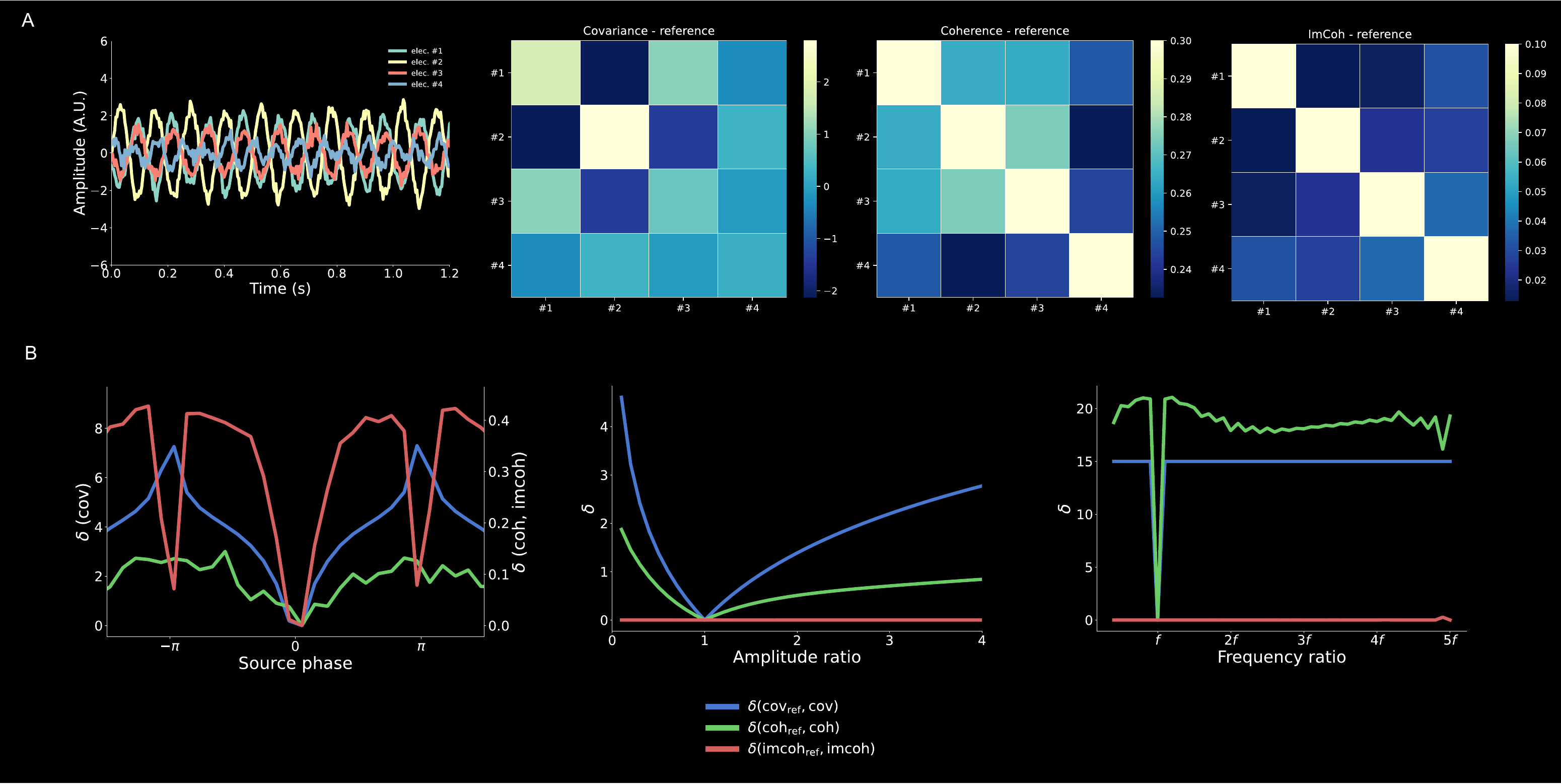}
		\caption{Toy Model. (A) On the left, synthetic EEG signals from four virtual electrodes. On the right, the associated covariance (Cov), coherence (Coh), and imaginary part of coherence (ImCoh) matrices used as references. (B) Influence of the key-parameters of one of the sources: the phase (on the left), the amplitude (center), and the frequency (on the right).
		\label{Figure18}%
		}
\end{figure*}
\clearpage

\begin{figure}
\centering
\includegraphics[width=\textwidth]{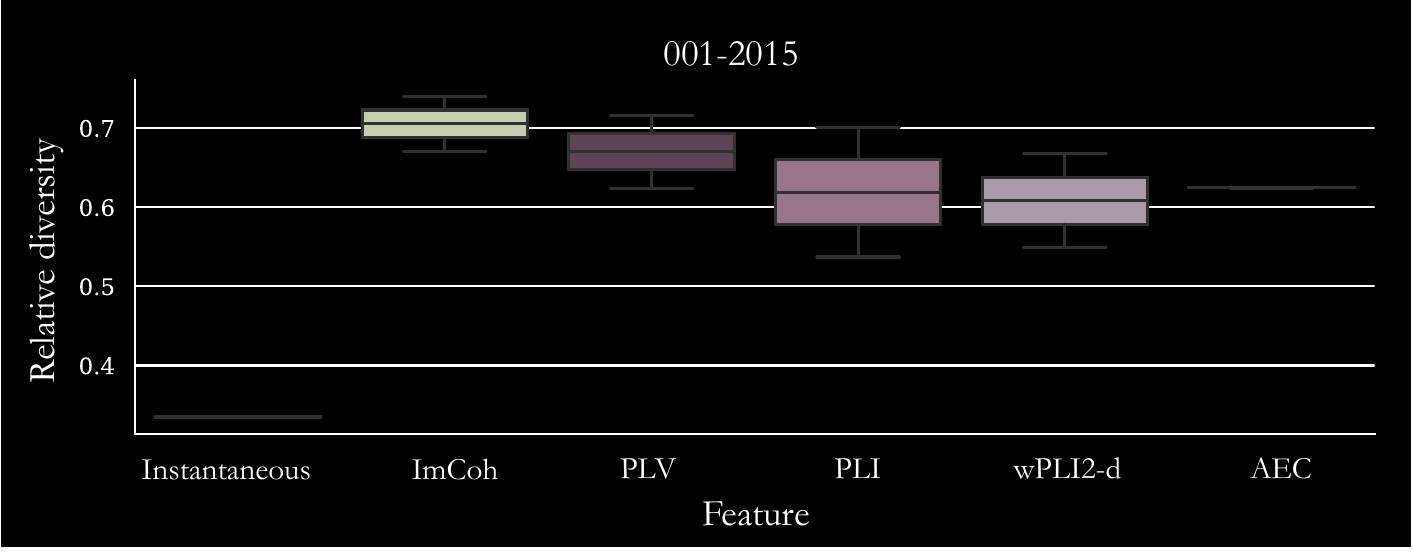}
\caption{Relative diversity plots, displayed for instantaneous coherence, ImCoh, PLV, PLI, wPLI debiased and AEC estimators compared to covariance. The results are obtained on the dataset BNCI001-2015.
}\label{Figure19}
\end{figure}

\begin{figure}
\centering
\includegraphics[width=\textwidth]{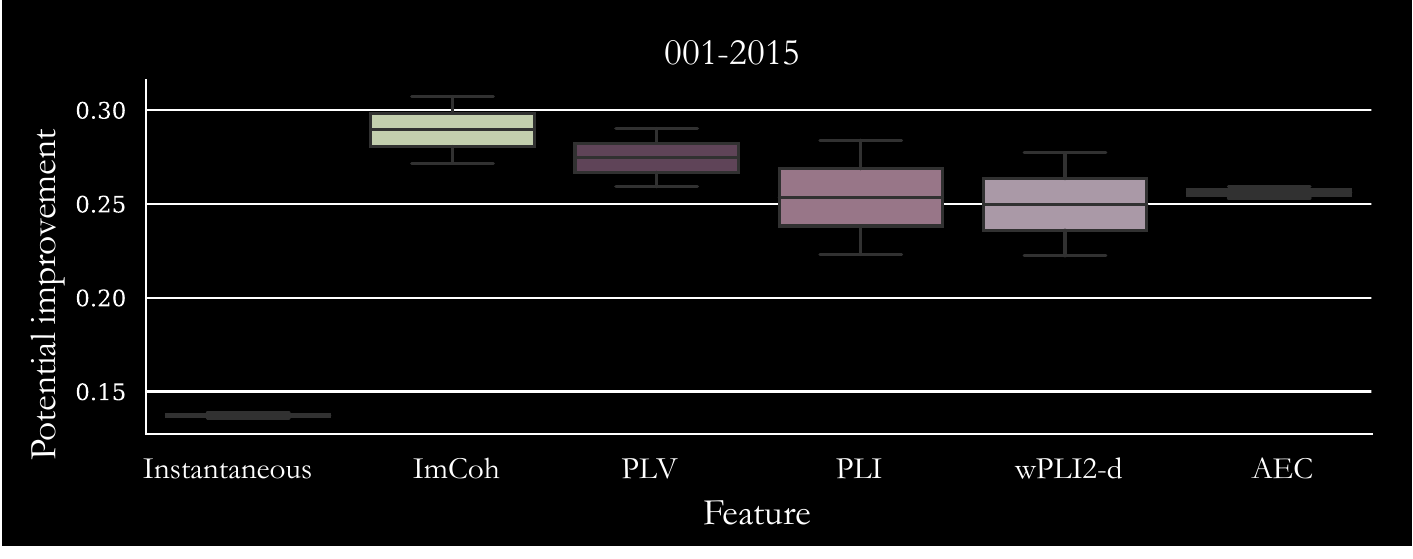}
\caption{Potential improvement plots displayed for the same estimators compared to covariance. The results are obtained on the dataset BNCI001-2015.
}\label{Figure20}
\end{figure}
\newpage
\clearpage